\documentclass[12pt]{article}
\pdfoutput=1
\usepackage{epsf,amsfonts,amssymb,epsfig,amsmath,graphics,slashed}
\usepackage{hyperref,graphicx, subfig,hep}
\newcommand{\nn}{\notag \\}

\begin{document}

\makeatletter
\renewcommand{\theequation}{\thesection.\arabic{equation}}
\@addtoreset{equation}{section}
\makeatother

\baselineskip 18pt

\begin{titlepage}

\vfill

\begin{flushright}
Imperial/TP/2012/JG/04\\
\end{flushright}

\vfill

\begin{center}
   \baselineskip=16pt
   {\Large\bf Supersymmetric quantum criticality\\
 supported by baryonic charges}
  \vskip 1.5cm
      Aristomenis Donos and Jerome P. Gauntlett\\
   \vskip .6cm
      \begin{small}
      \textit{Blackett Laboratory, 
        Imperial College\\ London, SW7 2AZ, U.K.}
        \end{small}\\*[.6cm]

\end{center}

\vfill

\begin{center}
\textbf{Abstract}
\end{center}

\begin{quote}
In the context of the $AdS_4\times Q^{111}$ solution of $D=11$ supergravity we construct 
supersymmetric zero temperature black brane solutions that interpolate between $AdS_4$ in the UV and $AdS_2\times \mathbb{R}^2$ in the IR.
The dual $N=2$ SCFT has a $U(1)^2$ baryonic symmetry and the solutions carry electric charge with respect to one of the $U(1)$ factors and
magnetic charge with respect to the other. The solutions
describe stable zero temperature ground states of the deformed SCFT which have finite entropy density. 
We also construct analogous supersymmetric solutions that flow to $AdS_2\times S^2$ and to $AdS_2\times H^2/\Gamma$ in the IR 
which, in addition, carry magnetic $R$-symmetry charge similar to other known wrapped brane solutions.
\end{quote}

\vfill

\end{titlepage}
\setcounter{equation}{0}


\section{Introduction}

The AdS/CFT correspondence provides an important forum for investigating the properties of strongly coupled
matter when held at finite chemical potential with respect to a conserved charge. It is of particular interest to examine
this issue in the context of ``top-down'' solutions of $D=10,11$ supergravity with well defined dual CFTs.

One very broad set-up is to consider the most general class of $AdS_4\times M_{6,7}$ solutions of $D=10,11$ supergravity which are dual to 
$N=2$ SCFTs in $d=3$ spacetime dimensions. For $D=11$ this includes the $AdS_4\times SE_7$ class of solutions, where  $SE_7$ is a seven-dimensional
Sasaki-Einstein space, as well as the general class of $AdS_4\times M_7$ solutions of \cite{Gabella:2012rc}, extending those of \cite{Gauntlett:2006ux}.
The internal space $M_{6,7}$ is either known to have, or is expected to have, an abelian isometry which is dual to the
abelian global $R$-symmetry of the dual SCFT.  
After Kaluza-Klein reduction on $M_{6,7}$ the isometry gives rise to a $D=4$ gauge-field
which is dual to the $R$-symmetry current. It is also expected \cite{Gauntlett:2007ma}
that there is a consistent Kaluza-Klein reduction from $D=11$ on $M_7$
or $D=10$ on $M_6$ which just keeps this gauge field and the metric leading to 
$D=4$ Einstein-Maxwell theory, the bosonic part of $D=4$ $N=2$ minimal gauged supergravity.
In fact, this has been explicitly proven in \cite{Gauntlett:2007ma} for the $SE_7$ class and the class of $M_7$ found in \cite{Gauntlett:2006ux}.
The consistency of the KK reduction means that any solution of the
Einstein-Maxwell theory uplifts to an infinite, universal class of exact solutions of $D=10,11$ supergravity. In particular, the uplifted electrically charged AdS-RN black brane solution
governs the high temperature behaviour of the entire class of dual $N=2$ SCFTs in $d=3$
when held at finite chemical potential with respect to the abelian $R$-symmetry.

A fascinating feature of the $D=4$ AdS-RN black brane solution is that it has non-zero entropy density at zero temperature, 
interpolating between $AdS_4$
in the UV and $AdS_2\times \mathbb{R}^2$ in the IR. If this solution describes the zero temperature physics, then the long-wavelength limit of the ground state is 
a locally quantum critical point dual to the $AdS_2\times \mathbb{R}^2$ solution. It is particularly interesting that such ground states can exhibit non Fermi-liquid behaviour \cite{Lee:2008xf,Liu:2009dm,Cubrovic:2009ye,Faulkner:2009wj} with the novel scaling near the Fermi surface being governed by the 
$AdS_2\times \mathbb{R}^2$ solution. However, in many situations the uplifted AdS-RN black brane solution is known to be 
unstable and hence cannot describe the
zero temperature ground states. In such cases the AdS-RN black brane solution becomes unstable at some finite temperature corresponding to the existence
of a new branch of black hole solutions which are dual to a new phase of the SCFT. Examples of instabilities of the
$D=4$ electrically charged AdS-RN black brane include superconducting phases 
\cite{Gubser:2008px,Hartnoll:2008vx,Hartnoll:2008kx,Denef:2009tp,Gauntlett:2009dn} 
as well as spatially modulated phases \cite{Donos:2011bh}.

It remains an interesting open question whether there are {\it any} top-down settings where the zero temperature limit of the AdS-RN black brane
is free from instabilities, and hence could provide a candidate ground state. For specific cases, such as when the internal space $M_{6}$ or $M_{7}$ is a homogeneous space,
it might be possible to address this issue at the perturbative level by analysing the full Kaluza-Klein spectrum. However, in addition one would also 
need to show that the solutions are free from non-perturbative instabilities which might be even more difficult to achieve.

One is motivated, therefore, to construct supersymmetric domain wall solutions of $D=10,11$ supergravity,
interpolating between $AdS_4$ in the UV and $AdS_2\times \mathbb{R}^2$ in the IR, with the supersymmetry guaranteeing the stability of the solution as a zero temperature  quantum critical ground state\footnote{Note that it should always be possible to heat up such
domain wall solutions, at least for small temperatures, simply because it corresponds to an irrelevant deformation in the
IR.}. 
In the context of the maximally supersymmetric $AdS_4\times S^7$ solution of $D=11$ supergravity, such supersymmetric solutions were
constructed using the $D=4$ $U(1)^4\subset SO(8)$ gauged supergravity in \cite{Almuhairi:2011ws,Donos:2011pn,Almheiri:2011cb}. However, these solutions are supported by purely magnetic 
charges\footnote{It is worth recalling, though, that there are two different boundary conditions for massless gauge-fields in $AdS_4$, corresponding
to two different CFTs. The two cases are related, essentially, by interchanging electric and magnetic charges \cite{Witten:2003ya}.}
with respect to the $U(1)^4$ global symmetry. In this paper we shall construct  similar supersymmetric domain wall solutions, 
which carry both electric and magnetic charges.

One strategy to construct such a domain wall solution is as follows. 
First, find an appropriate supersymmetric solution of $D=10,11$ supergravity that
contains an $AdS_2\times\mathbb{R}^2$ factor. Second, find a suitable supersymmetric 
solution of $D=10,11$ supergravity with an $AdS_4$ factor that could provide the UV asymptotics of the domain
wall. Third, construct the domain wall solution. For the first step we can look amongst the rich family of supersymmetric solutions of $D=11$ supergravity found in
\cite{Gauntlett:2006ns,Donos:2008ug}, building on the elegant classification of \cite{Kim:2006qu}. More specifically, we will
focus on a family of supersymmetric $AdS_2\times \mathbb{R}^2\times S^2\times S^2\times S^3$ solutions of $D=11$ supergravity constructed in \cite{Donos:2008ug}. The topology of this solution suggests that a candidate supersymmetric solution for
the second step is the $AdS_4\times Q^{111}$ solution of $D=11$ supergravity \cite{D'Auria:1983vy}.

Indeed, all of our new supersymmetric solutions are in the context of the $AdS_4\times Q^{111}$ solution, 
or various orbifolds thereof, for which there has been recent progress on elucidating the dual $N=2$ SCFTs \cite{Franco:2008um,Franco:2009sp,Klebanov:2010tj,Benishti:2010jn} (for older work see \cite{Fabbri:1999hw}). Recall that
$Q^{111}$ is a seven-dimensional Sasaki-Einstein manifold that is a $U(1)_R$ fibration over $S^2\times S^2\times S^2$
with topology $S^2\times S^2\times S^3$. The $AdS_4\times Q^{111}$ solution arises
after placing M2-branes at the tip of the Calabi-Yau four-fold cone over $Q^{111}$ and taking the near horizon limit.
The isometry group of $Q^{111}$
is $SU(2)^3\times U(1)_R$, with the $U(1)_R$ factor generated by the Reeb Killing vector.
A Kaluza-Klein reduction on $Q^{111}$ leads to an $N=2$ $D=4$ gauged supergravity theory. 
The $U(1)_R$ isometry gives rise to a gauge field that
lives in the $N=2$ graviton multiplet, while the
$SU(2)^3$ isometries will give rise to vector multiplets, which will play no role in our solutions.
Since the second Betti number of $Q^{111}$ is two, there will also be $U(1)^2$ baryonic symmetry. Indeed
using the two independent harmonic two-forms on $Q^{111}$
the dimensional reduction of the three-form potential of $D=11$ supergravity will give rise
to two baryonic gauge-fields each living in a Betti vector multiplet \cite{Fabbri:1999hw}. Recall that
none of the Kaluza-Klein spectrum is charged under the baryonic $U(1)^2$ symmetry
and that the only states that can carry such charges are wrapped M2-branes or
M5-branes (see \cite{Klebanov:2010tj,Benishti:2010jn} for a more precise discussion).

In \cite{Klebanov:2010tj} zero temperature black brane solutions, without supersymmetry, that interpolate between 
a deformation of $AdS_4\times Q^{111}$ in the UV and a new $AdS_2\times\mathbb{R}^2$ solution in the IR were constructed which carry 
electric charge\footnote{The ambiguity \cite{Witten:2003ya}
for the $AdS_4$ boundary conditions that one imposes for these baryonic $U(1)^2$ gauge-fields is discussed in the present context in 
\cite{Klebanov:2010tj,Benishti:2010jn}.
For definiteness, we are adopting a language in which the baryonic gauge-fields arising from the $D=11$ three-form potential using the harmonic two-forms
have standard boundary conditions. Specifically, wrapped M2-branes and M5-branes carry electric and magnetic charges, respectively (in contrast to \cite{Benishti:2010jn}).} 
with respect to one of the $U(1)^2$ baryonic symmetries. Here, we will construct supersymmetric solutions
that  carry electric charge with respect to one of the baryonic symmetries and magnetic charge with respect to the other. The solutions interpolate between a deformation of
$AdS_4\times Q^{111}$ in the UV with the class of supersymmetric $AdS_2\times \mathbb{R}^2\times S^2\times S^2\times S^3$ solutions constructed in \cite{Donos:2008ug}  in the IR.

A simple extension involves replacing the $\mathbb{R}^2$ factor with an $S^2$ or an $H^2$ factor, obtaining 
supersymmetric solutions interpolating between $AdS_4$ in the UV and a class of $AdS_2\times S^2$ or $AdS_2\times H^2$ fixed points in the IR, again first constructed
in \cite{Donos:2008ug}.
It is also possible to take a quotient of $H^2$ to obtain a compact Riemann surface of genus greater than one while preserving supersymmetry.
The domain wall solutions with $S^2$ and $H^2$ factors share some similarities with known supersymmetric solutions describing branes wrapping supersymmetric cycles.
In particular, the solutions asymptote in the UV to $AdS_4$ in Poincar\'e-type coordinates with the three-dimensional slices at constant radius
having topology $\mathbb{R}\times S^2$ or $\mathbb{R}\times H^2$, respectively. 
This is precisely what happens in wrapped brane solutions \cite{Maldacena:2000mw} (see \cite{Gauntlett:2003di} for a review and \cite{Anderson:2011cz,Bah:2012dg}
for more recent work). Furthermore,
in addition to the baryonic charges our new solutions also carry magnetic $R$-symmetry charges as in the known wrapped brane solutions. It is interesting that
in contrast to the wrapped
brane solutions in \cite{Maldacena:2000mw,Gauntlett:2001qs}, where only $H^2$ factors are allowed, here the baryonic charges also allow $S^2$ factors.
It is likely that our solutions can also be constructed in a $D=4$ gauged supergravity arising from 
a consistent KK reduction of $D=11$ supergravity on $Q^{111}$, extending \cite{Gauntlett:2009zw}. 
From this point of view our new solutions share some similarities with
supersymmetric zero temperature $AdS_4$ black hole solutions that have been constructed in $D=4$ gauged supergravity in
\cite{Cacciatori:2009iz,Dall'Agata:2010gj,Hristov:2010ri,Barisch:2011ui}.

In section 2 we discuss the supersymmetric black branes interpolating between $AdS_4$ and $AdS_2\times\mathbb{R}^2$.
The solutions interpolating between $AdS_4$ and $AdS_2\times S^2$ or $AdS_4\times H^2$ are discussed in section 3 and we
conclude with some final comments in section 4. The paper contains three appendices.

\section{The supersymmetric flow from $AdS_{4}$ to $AdS_2\times \mathbb{R}^2$}

We will construct bosonic solutions of $D=11$ supergravity using the conventions of \cite{Gauntlett:2002fz}.
In particular the Bianchi identity and the equation of motion for the four-form are given by
\begin{equation}\label{eq:4form_eom}
dF=0,\qquad d*F+\frac{1}{2}\,F\wedge F=0\,.
\end{equation}
Furthermore, a Killing spinor satisfies
\begin{align}\label{eq:gravitinovariation}
\nabla_M\varepsilon+\frac{1}{24}\left(3\slashed F\,\Gamma_{M} -\Gamma_{M}\,\slashed F\right)\varepsilon=0\,,
\end{align}
where $\slashed F=\tfrac{1}{4!}F_{ABCD}\Gamma^{ABCD}$ and $\varepsilon$ is a $D=11$ Majorana spinor. The $D=11$ gamma-matrices satisfy
$\Gamma_{012345678910}=1$.

\subsection{The $AdS_4\times Q^{111}$ solution}

We begin by recalling the supersymmetric $AdS_4\times Q^{111}$ solution.
The metric and four-form are given by
\begin{align}\label{Q111}
ds^{2}=&e^{4\rho}(-dt^2+dx_1^2+dx_2^2)+d\rho^2+ds_{1}^{2}+ds_{2}^{2} +ds_{3}^{2}+\eta^{2}\,,\notag\\
F=&6e^{6\rho}dt\wedge dx_1\wedge dx_2\wedge d\rho\,,
\end{align}
where 
\begin{align}\label{spherenorm}
ds_{i}^{2}=\tfrac{1}{8}\left(d\theta_{i}^{2}+\sin^{2}\theta_{i}\,d\phi_{i}^{2} \right),\qquad
\eta=\tfrac{1}{4}(d\psi+P_1+P_2+P_3)\,,
\end{align}
with $dP_i=\tilde{\mathrm{vol}}_i\equiv \sin\theta_{i}\,d\theta_{i}\wedge d\phi_{i}$. For $Q^{111}$ the period of $\psi$ is given by $\psi\cong\psi+4\pi$.
The solution, as well as various orbifolds thereof, preserves four Poincar\'e and four superconformal supersymmetries. The
corresponding dual $d=3$ $N=2$ SCFTs have been discussed in \cite{Franco:2008um,Franco:2009sp,Klebanov:2010tj,Benishti:2010jn}.

It is helpful to record the explicit form of the Poincar\'e Killing supersymmetries. 
In the obvious orthonormal frame (see \eqref{elf} below) 
the Poincar\'e supersymmetries satisfy the algebraic conditions
\begin{align}
\label{eq:projections}
&\Gamma^{4567}\varepsilon=-\varepsilon,\quad \Gamma^{4589}\varepsilon=-\varepsilon,\quad\Gamma^{453\sharp}\varepsilon=-\varepsilon
\nn
&\Rightarrow\Gamma^{012}\,\varepsilon=-\varepsilon\,.
\end{align}
The first line corresponds to the projections associated with the Calabi-Yau four-fold cone over the Sasaki-Einstein space $Q^{111}$ while
the second line corresponds to the fact that we can place a membrane at the apex of this cone without breaking further supersymmetry. Furthermore,
we have $\varepsilon=e^{\rho}\varepsilon_0$ where $\varepsilon_0$ only depends on the coordinates of $Q^{111}$ and satisfies
\begin{align}\label{sasein}
\hat{\nabla}_{m}\varepsilon_0-\frac{1}{2}\Gamma^3\Gamma_{m}\varepsilon_0=0\,,
\end{align}
where $\hat\nabla$ is the Levi-Civita connection on $Q^{111}$ with coordinates $y^m$.

\subsection{The flow equations}
We aim to construct a supersymmetric flow from this $AdS_4$ solution to an $AdS_2\times \mathbb{R}^2$ solution after switching on suitable
deformations. The ansatz we shall consider is given by
\begin{align}\label{eq:11d_ansatz}
ds^{2}=& -e^{2A}dt^{2}+e^{2B}\,\left(dx_{1}^{2}+dx_{2}^{2}\right)+d\rho^{2}+e^{2U_{1}}\,\left(ds_{1}^{2}+ds_{2}^{2} \right)+
e^{2U_{3}}\,ds_{3}^{2}+e^{2V}\,\eta^{2}\,,\notag\\
F=&Z\,dt\wedge d\rho\wedge dx_{1}\wedge dx_{2}+
dt\wedge d\rho\wedge\left(g_1J_{1}+g_{1}J_{2}+g_{3}J_{3} \right)\notag\\
&+\lambda\,dx_{1}\wedge dx_{2}\wedge\left(J_{1}-J_{2} \right)+d\left[f \eta\wedge\left(J_{1}-J_{2}\right)\right]\,,
\end{align}
where we have used the K\"ahler forms $J_i$ for the two-spheres defined by
\begin{align}
J_i=\tfrac{1}{8}\mathrm{vol}_i,\qquad i=1,2,3\,.
\end{align}
Notice that $d\eta=2(J_1+J_2+J_3)$.
Furthermore, $\lambda$ is a constant, and $A, B, U_1, U_3, V, Z, g_1, g_3$ and $f$ are functions that depend on the radial coordinate $\rho$ only.
One can easily generalise this ansatz including the introduction of obvious functions $U_2, g_2$; we will return to this at the end of the paper.

Clearly the $AdS_4\times Q^{111}$ solution is recovered via
\begin{align}\label{adsq111}
&A= B=2\rho,\qquad Z=6e^{6\rho}\,,\nn
&U_{1}=U_{3}=V=f=g_1=g_3=\lambda=0\,.
\end{align}
The ansatz automatically solves the Bianchi identity for the four-form in \eqref{eq:4form_eom}.
We next impose the four-form equation of motion given in \eqref{eq:4form_eom}, deducing that
\begin{align}\label{eq:4form_sol}
Z=&\left(\alpha-2f^{2}\right)\,e^{A+2B-V-4U_{1}-2U_{3}}\,,\notag\\
g_{1}=&\beta\,e^{A-2B-V-2U_{3}}\,,\notag\\
g_{3}=&-2\left(\beta+\lambda\,f\right)\,e^{A-2B-V-4U_{1}+2U_{3}},
\end{align}
where $\alpha$ and $\beta$ are constants. It also
yields the second order equation of motion
\begin{equation}\label{eq:f_so_eom}
-\left(Ge^{-2V-4U_{1}} f^{\prime}\right)^{\prime}+4fGe^{-4U_{3}-4U_{1}}=2fZ+\lambda g_{3}\,,
\end{equation}
where $G=e^{A+2B+V+4U_{1}+2U_{3}}$.
We will see later that this equation is implied by the Killing spinor conditions.
We take $\alpha=6$ to make contact with the $AdS_{4}\times Q^{111}$ solution \eqref{Q111}.  
As we will shall discuss in section \ref{sec:dual}, the constants $\lambda$ and $\beta$ will correspond
to deformations of the $AdS_{4}\times Q^{111}$ solution that drive the RG flow in the domain wall solution. 
Note that when $\lambda\ne 0$ it can be set to any convenient value by scaling $x_{i}$ and shifting $B$.

\subsection{Killing spinor analysis}
We now define the elfbein
\begin{align}\label{elf}
e^{0}&=e^{A}\,dt,\qquad e^{1}=e^{B}\,dx_{1},\qquad e^{2}=e^{B}\,dx_{2},\qquad e^{3}=d\rho\,,\notag\\
e^4&=\tfrac{e^{U_1}}{2\sqrt 2}d\theta_1,\qquad e^5=\tfrac{e^{U_1}}{2\sqrt 2}\sin\theta_1 d\phi_1,\qquad
e^6=\tfrac{e^{U_1}}{2\sqrt 2}d\theta_2,\qquad e^7=\tfrac{e^{U_1}}{2\sqrt 2}\sin\theta_2 d\phi_2\,,
\nn
e^8&=\tfrac{e^{U_3}}{2\sqrt 2}d\theta_3,\qquad e^9=\tfrac{e^{U_3}}{2\sqrt 2}\sin\theta_3 d\phi_3,\qquad
e^{\sharp}=e^{V}\,\eta,
\end{align}
and assume that $\varepsilon$ has the form
\begin{equation}
\varepsilon=e^{C}\,\varepsilon_0\,,
\end{equation}
where $C$ is a function of $\rho$ and $\varepsilon_0$ is independent of $\rho$, $t$ and $x_{i}$.
We continue to impose the projection conditions \eqref{eq:projections} and in addition we demand that
\begin{align}\label{projs2}
\Gamma^{1245}\varepsilon=-\varepsilon\,.
\end{align}
We also assume that $\varepsilon_0$ satisfies the differential condition for the Killing spinors on $Q^{111}$ given
in \eqref{eq:projections}.
We then find that the $D=11$ Killing spinor equations lead to $C=A/2$ and
the following system of first order differential equations
\begin{align}\label{eq:fo_system}
A^{\prime}-\frac{e^{-A}}{3}\,\left(e^{-2B}Z+2g_{1}e^{-2U_{1}}+g_{3}e^{-2U_{3}} \right)&=0\notag\\
B^{\prime}-\frac{e^{-A}}{6}\,\left(2e^{-2B}Z-2g_{1}e^{-2U_{1}}-g_{3}e^{-2U_{3}} \right)&=0\nn
U_{1}^{\prime}-e^{V-2U_{1}}+\frac{e^{-A}}{6}\left(e^{-2B}Z-g_{1}e^{-2U_{1}}+g_{3}e^{-2U_{3}}\right)&=0\notag\\
U_{3}^{\prime}-e^{V-2U_{3}}+\frac{e^{-A}}{6}\left(e^{-2B}Z+2g_{1}e^{-2U_{1}}-2g_{3}e^{-2U_{3}} \right)&=0\notag\\
V^{\prime}+2e^{V-2U_{1}}+e^{V-2U_3}
-4e^{-V}
+\frac{e^{-A}}{6}\left(e^{-2B}Z+2g_{1}e^{-2U_{1}}+g_{3}e^{-2U_{3}}  \right)&=0\notag\\
f^{\prime}+2e^{V-2U_{3}}f+\lambda\,e^{V-2B}&=0
\end{align}
(for more details see appendix \ref{killspinan}).
One can now show that the second order equation \eqref{eq:f_so_eom} is automatically implied by these equations.
Since we have satisfied the Bianchi identity and the equation of motion for the four-form, we can use a result of
\cite{Gauntlett:2002fz} to deduce
that any solution to the differential equations \eqref{eq:fo_system} will give rise to a supersymmetric solution
of $D=11$ supergravity preserving at least two supersymmetries.

\subsection{The $AdS_{2}\times\mathbb{R}^2$ fixed points}\label{sec:AdS2}
In addition to the $AdS_4$ solution \eqref{adsq111}, the differential equations \eqref{eq:fo_system} also admit,
for $\lambda,\beta\ne 0$, a one-parameter family of supersymmetric $AdS_{2}\times\mathbb{R}^2$ fixed points, parametrised by $\beta$, first constructed
in \cite{Donos:2008ug}. We present the general family of solutions in appendix \ref{sec:AdS_familyB}. To keep the discussion simple we just 
record the particular solution that has $U_1=U_3$ given by
\begin{align}\label{fptsol}
A&=2^{11/6}{3^{1/3}}\,\rho,\quad B=-\frac{1}{2}\ln\left(\frac{2\cdot 6^{1/6}}{\lambda} \right),\quad \beta=\sqrt{\frac{2}{3}}\,\lambda,\nn
U_{1}&=U_{3}=-\frac{1}{6}\,\ln\left(\frac{4}{3}\right),\quad V=\frac{1}{3}\ln\left(\frac{2^{1/2}}{3} \right),\quad f=-\sqrt{\frac{3}{2}},
\end{align}
Choosing $\lambda=8\sqrt{3}$ for convenience, the resulting $D=11$ solution can be then written in the form
\begin{align}\label{uplift2}
ds^{2}=& \frac{1}{a^2}\left[ds^2(AdS_2)+a^3\left(dx_{1}^{2}+dx_{2}^{2}+\tfrac{1}{l_1}\left(d\tilde s_{1}^{2}+d\tilde s_{2}^{2} +d\tilde s_{3}^{2}\right)\right)+
(d\psi+P)^2\right]\,,\notag\\
F=&\mathrm{vol}(AdS_2)\wedge \left[dx_{1}\wedge dx_{2}+\frac{2}{3l_1}\left(\tilde{\mathrm{vol}}_1+\tilde{\mathrm{vol}}_2+\tilde{\mathrm{vol}}_3\right)\right]\nn
&+2m_{13}\left[-dx_1\wedge dx_2\wedge\frac{1}{l_1}(\tilde{\mathrm{vol}}_1-\tilde{\mathrm{vol_2}})+\frac{1}{l_1^2}\tilde{\mathrm{vol}}_3\wedge(\tilde{\mathrm{vol}}_1-\tilde{\mathrm{vol}}_2)\right]\,,
\end{align}
with $P=P_1+P_2+P_3$, $a=2^{1/6} 3^{1/3}$, $l_1=2^{11/2}$ and $m_{13}=-\sqrt{3}l_1/2$. Furthermore,
$ds^2(AdS_2)$ and $\mathrm{vol}(AdS_2)$ are the metric and volume-form on a unit radius $AdS_2$.
This agrees\footnote{
In \cite{Donos:2008ug} we should set
$l_{1}=l_{2}=l_{3}$, $l_4=0$, $J_{1}=dx_{1}\wedge dx_{2}$, $m_{12}=0$ and $m_{13}=-m_{14}$.
We also point out that there are two typos in the first of the two equations (3.21) which can be fixed by dividing the expression for $F_{2}$ by a factor of $2$ and replacing $l_1$ in the first term on the right hand side with $l_4$.} with
the solutions of section (3.2) in \cite{Donos:2008ug}. Observe that the topology of the internal space is the same as that of $Q^{111}$, namely 
$S^2\times S^2\times S^3$.

\subsection{Supersymmetric domain walls}

We now construct, numerically, a flow from the $AdS_4\times Q^{111}$ solution \eqref{Q111}
to the $AdS_2\times \mathbb{R}^2 \times S^2\times S^2\times S^3$ solution \eqref{uplift2}. As usual we develop a series
expansion of the differential equations \eqref{eq:fo_system}
about both the $AdS_4$ UV fixed point \eqref{adsq111} and the $AdS_2$ IR fixed point \eqref{fptsol} and then use a shooting technique
to match them. In order to hit the fixed point \eqref{fptsol} in the IR we now set 
$\beta=\sqrt{\frac{2}{3}}\,\lambda$.

In constructing an expansion about the $AdS_4$ UV fixed point \eqref{adsq111},
 there are three modes of interest corresponding to deformations by relevant operators in the dual SCFT
(discussed in section \ref{sec:dual}).
Indeed we have the following perturbative modes: 
\begin{align}\label{modesone}
(\delta U_1,\delta U_3,\delta V)&\sim 
(1,1,-6)e^{-2\Delta_1\rho}\,,\nn
(\delta U_1,\delta U_3)&\sim 
(1,-2)e^{-2\Delta_2\rho}\,,\nn
\delta f&\sim 
e^{-2\Delta_2\rho}\,,
\end{align}
where $\Delta_{1}=4$ and $\Delta_{2}=1$. 
We can then develop an expansion about the deformed $AdS_4$ solution in terms of the deformation parameter 
$\lambda$
(with $\beta=\sqrt{\frac{2}{3}}\,\lambda$) specified by three constants $c_1,c_2$ and $ c_3$
which will be constants of integration for our boundary value problem. In more detail, as $\rho\to\infty$ we have
\begin{align}\label{uvexpsads4}
A&=2 \rho+\frac{\lambda ^2}{20} \left(-12 c_{2}^2+c_{3}^2\right) e^{-4 \rho} +\frac{\lambda ^2}{27}   \left(4 c_{3}-24 c_{2}^3 \lambda +c_{2} \left(-4 \sqrt{6}+10 c_{3}^2 \lambda \right)\right)e^{-6 \rho}+\cdots\notag\\
B&=2 \rho+\frac{\lambda ^2}{20} \left(-12 c_{2}^2+c_{3}^2\right) e^{-4 \rho} -\frac{\lambda ^2}{54}   \left(c_{3}+48 c_{2}^3 \lambda -10 c_{2} \left(\sqrt{6}+2 c_{3}^2 \lambda \right)\right)e^{-6 \rho}+\cdots\nn
U_{1}&=\lambda c_{2} e^{-2 \rho}  +\frac{\lambda ^2}{60} \left(84 c_{2}^2-2 c_{3}^2-\frac{5 \sqrt{6}}{\lambda }\right)e^{-4 \rho}  +\dots+\lambda ^4c_{1} e^{-8 \rho} +\cdots\notag\\
U_{3}&=-2\lambda c_{2} e^{-2 \rho}  + \lambda ^2\left(-\frac{8 c_{2}^2}{5}-\frac{c_{3}^2}{30}+\frac{1}{\sqrt{6} \lambda }\right)e^{-4 \rho} +\cdots+\lambda ^4(c_{1}+\dots) e^{-8 \rho} +\cdots\notag\\
V&= \lambda ^2\left(-\frac{18 c_{2}^2}{5}-\frac{c_{3}^2}{30}\right) e^{-4 \rho}+\dots+\lambda^4(-6c_1+\cdots)e^{-8\rho}+\cdots\nn
f&=\lambda c_{3} e^{-2 \rho} +\lambda ^2\left(4 c_{2} c_{3}+\frac{1}{2 \lambda }\right)e^{-4 \rho} \cdots
\end{align}
where, for simplicity of presentation, we have suppressed a few order $e^{-6\rho}$ terms, but have included all terms at order $e^{-8\rho}$ that are linear in the
$c_i$. Notice that the terms linear in the $c_i$ correspond to the three modes in \eqref{modesone}.

We next develop an expansion about the $AdS_2\times\mathbb{R}^2$ IR fixed point \eqref{fptsol}. We find that there are two modes that are active, corresponding
to two irrelevant operators in the SCFT dual to the $AdS_2$ fixed point, with scaling dimensions $\Delta_a=1+\delta_a$ with $\delta_{1}=1$ and $\delta_{2}\approx 1.48$.
The expansion is specified by three free constants $d_1,d_2$ and $d_3$, which are three further constants of integration in our boundary value problem.
Schematically, as $\rho\to -\infty$ we have
\begin{align}
A&=a \rho +d_{3}-6d_1  \,e^{a \rho\delta_{1}} -(1.30)d_2 \,e^{a \rho\delta_{2}} +\cdots\notag\\
B&=-\frac{1}{2}\log\left(\frac{2\ 6^{1/6}}{\lambda }\right)+\frac{10}{7} d_1 \,e^{a \rho\delta_{1}} -(7.58)\,d_2 e^{a \rho\delta_{2}} +\cdots\notag\\
U_{1}&=-\frac{1}{6}\log\left(\frac{4}{3}\right)+d_1 e^{a \rho\delta_{1}} +d_2 e^{a \rho\delta_{2}} +\cdots\notag\\
U_{3}&=-\frac{1}{6}\log\left(\frac{4}{3}\right)-\frac{2}{7}d_1 \,e^{a \rho\delta_{1}} +(5.02) d_2 \,e^{a \rho\delta_{2}} +\cdots\notag\\
V&=\log\left(\frac{2^{1/6}}{3^{1/3}}\right)+\frac{10}{7}d_1 \,e^{a \rho\delta_{1}}+(1.11)d_2 \,e^{a \rho\delta_{2}} +\cdots\notag\\
f&=-\sqrt{\frac{3}{2}}+\frac{3\sqrt{6}}{7}d_1 \,e^{a \rho\delta_{1}} -(5.66) d_2\,e^{a \rho\delta_{2}} +\cdots
\end{align}
with $a=2^{11/6}{3^{1/3}}$ and the numbers in parentheses are numerical approximations to coefficients that appear in the expansion.

\begin{figure}[t!]
\centering
\includegraphics[width=0.45\textwidth]{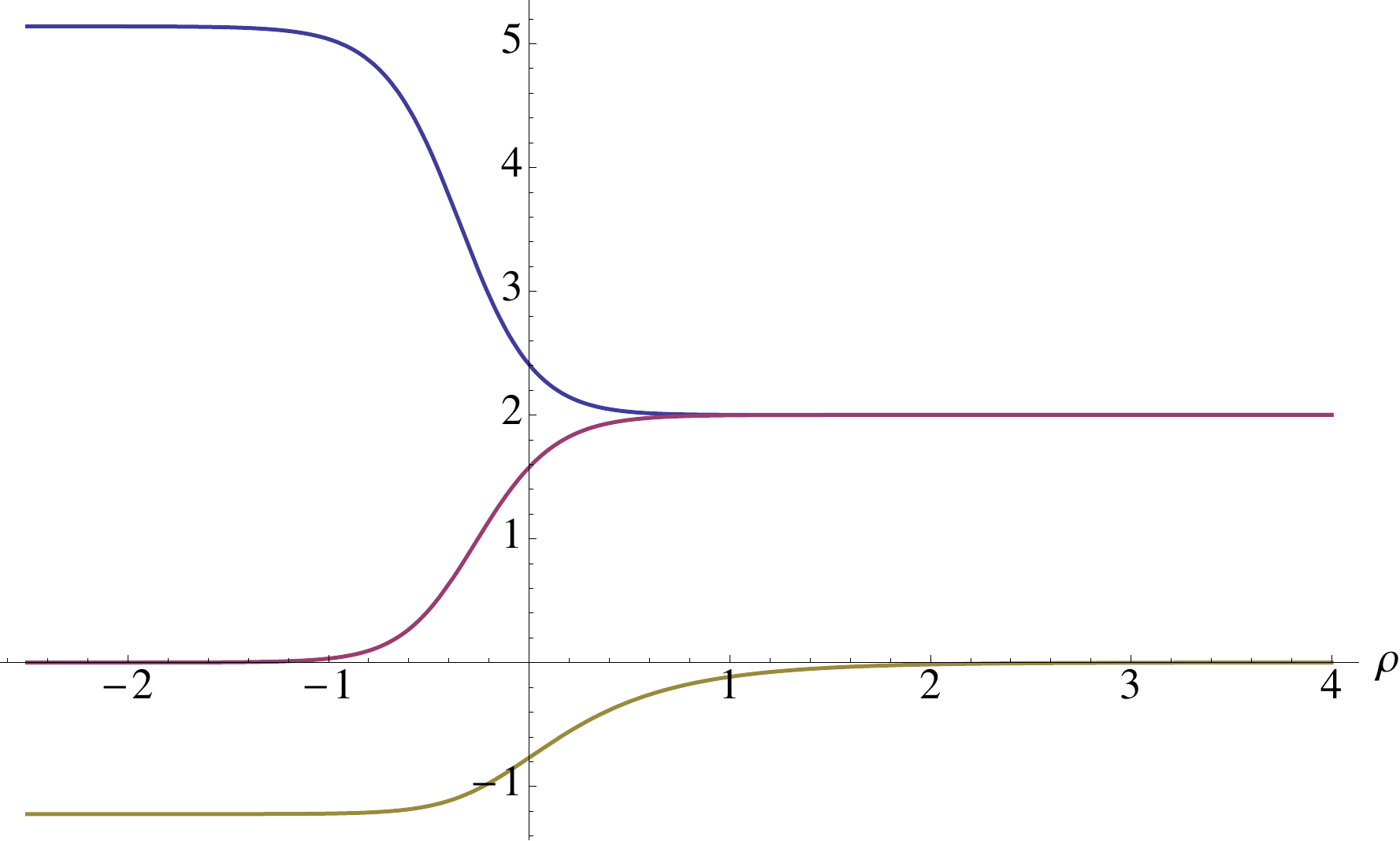}\qquad
\includegraphics[width=0.45\textwidth]{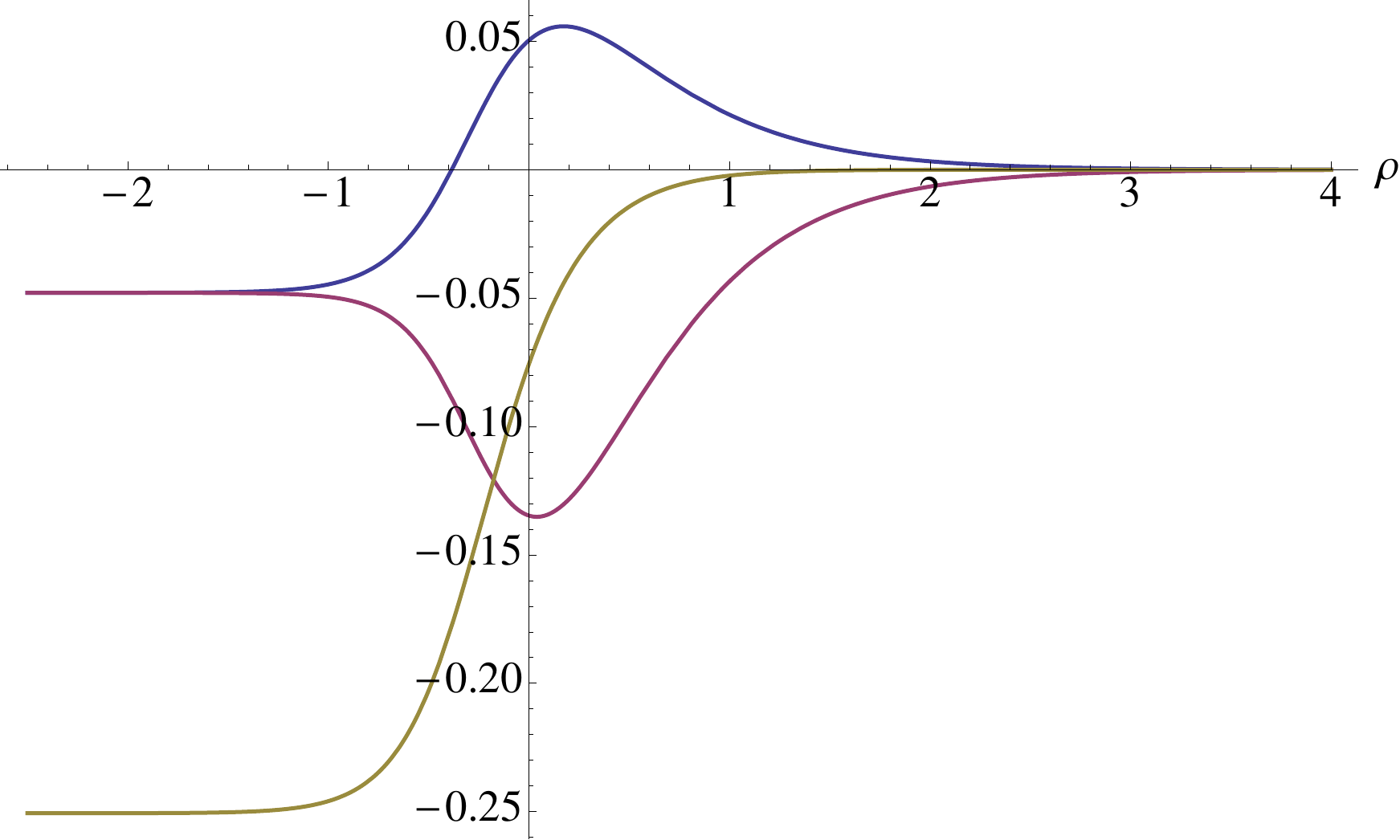}
\caption{Supersymmetric domain wall solutions interpolating between $AdS_4\times Q^{1,1}$ and $AdS_2\times \mathbb{R}^2\times S^2\times S^2\times S^3$ given
by \eqref{fptsol}.
In the left panel, top to bottom, we have plotted the functions $A'$ (blue), $B'$ (red), $f$ (green). In the right panel we have plotted, top to bottom from the left,
$U_1$ (blue), $U_3$ (red) and $V$ (green).
\label{fig:dwall1}}
\end{figure}

We thus have a system of six first order differential equations given in \eqref{eq:fo_system}
and a total of six constants of integration appearing in the UV and IR expansions.
Using a numerical shooting method we find the unique solution
\begin{align}\label{uvirvalues}
c_{1}=0.59\ldots,\qquad
c_2=-0.28\ldots,\qquad
c_{3}=1.12\ldots,\notag\\
d_{1}=-0.035\ldots,\qquad
d_{2}=0.18\ldots,\qquad
d_{3}=-0.84\ldots,
\end{align}
where without loss of generality we have set $\lambda=1$.
In figure \ref{fig:dwall1} we have plotted the behaviour of the functions appearing in the solution.
It is worth noting that even though both the $AdS_4$ and the $AdS_2\times\mathbb{R}^2$ fixed points
have $U_1=U_3$, the flow between them has $U_1\ne U_3$.

We anticipate that similar supersymmetric domain wall solutions exist interpolating between $AdS_4$ and the more general one-parameter family
of $AdS_2\times\mathbb{R}^2$ solutions presented in appendix \ref{sec:AdS_familyB}. In particular, we have checked numerically 
that the IR solutions always have two irrelevant operators that one can use to shoot out with, one with $\delta=1$ and the other with $\delta$ monotonically
increasing from $0$ to $1.5$ as the parameter $m$ labelling the solutions in \eqref{fullsetsol} is varied from $1$ to $\infty$. Thus there will again be three integration constants in the IR and three in the UV and so we expect to find a unique solution.
We have also constructed in detail the solutions for a couple of other cases.

It is straightforward to calculate the potential for a static M2-brane probe with world-volume $(t, x_1, x_2)$
in the domain wall geometry. As in section 4.1 of \cite{Herzog:2009gd} there are two contributions to the potential energy per
unit area as a function of the radial coordinate $\rho$. Up to a constant of proportionality there is a contribution
from the area of the M2-brane given by $v_g(\rho)=e^{A+2B}$ and a contribution from the coupling to the three-form flux given
by $v_e(\rho)'=\pm Z$. An analysis of the BPS equations \eqref{eq:fo_system}
reveals that $Z=e^{A+2B}$ and hence the potential will vanish for 
a M2-brane, as we expect by supersymmetry, and will be non-zero for an anti-M2-brane. An analysis of probe M2-branes wrapping other cycles will be 
left for future work.

\subsection{KK reduction and dual SCFT interpretation}\label{sec:dual}
Using the $AdS_4$ UV behaviour of the domain wall solution, specified by the constants $\lambda,\beta$ and $c_i$,
and our knowledge of the KK spectrum on $Q^{111}$,
we can draw some conclusions about the interpretation of these domain wall solutions in the dual $N=2$ SCFT. 

It is helpful to recall the consistent KK truncation of $D=11$ supergravity on an arbitrary $SE_7$ space presented in  \cite{Gauntlett:2009zw}.
The KK reduction leads to an $N=2$ $D=4$ gauged supergravity coupled to a vector multiplet and a hypermultiplet. Expanding about
the supersymmetric $AdS_4$ vacuum, these KK modes arrange themselves into $OSp(2|4)$ multiplets. There is a massless graviton multiplet 
which contains a massless vector dual to the $R$-symmetry current. There is also a long vector multiplet
containing a massive vector, dual to an operator of dimension $\Delta=5$, as well as five scalar fields, one of which (a squashing mode)
is dual to an operator of dimension $\Delta=4$. 

Since the second Betti number of $Q^{111}$ is two, there will be two ``Betti vector multiplets" in the KK spectrum corresponding to
$U(1)^2$ baryonic symmetry in the dual SCFT \cite{Fabbri:1999hw}.
It should be possible\footnote{The extension will be analogous to the extension of the KK reduction of type IIB on $T^{11}$
\cite{Cassani:2010na,Bena:2010pr}
from that on an arbitrary $SE_5$ \cite{Cassani:2010uw,Liu:2010pq,Gauntlett:2010vu,Skenderis:2010vz}.} to
extend the consistent KK reduction of \cite{Gauntlett:2009zw} to include the two 
$N=2$ Betti vector multiplets in the $D=4$ gauged supergravity theory, and moreover, our domain wall solutions should be solutions of this theory. Expanding about the $AdS_4$ vacuum, there will be two additional massless vector
multiplets of $OSp(2|4)$, containing two vectors, dual to the $U(1)^2$ baryonic currents, and also two scalars and two
pseudo-scalars of dimension $\Delta=1,2$.

To see how the vector fields in the Betti multiplets arise in the KK reduction, we
choose two linearly independent harmonic two-forms on $Q^{111}$
given by
\begin{align}
\omega^{(1)}&=\tfrac{1}{8}(\mathrm{vol}_1-\mathrm{vol}_2)\,,\nn
\omega^{(2)}&=\tfrac{1}{8}(\mathrm{vol}_1+\mathrm{vol}_2-2\mathrm{vol}_3)\,.
\end{align}
Indeed it is straightforward to see that these are both closed and co-closed on $Q^{111}$.
Notice that $\omega^{(2)}$ is a harmonic form which exists on any $SE_7$ which can be
written as a $U(1)_R$ fibration over the product of a K\"ahler-Einstein four-manifold with a two-sphere. On the other hand
the existence of the harmonic two-form $\omega^{(1)}$ arises because for $Q^{111}$ the K\"ahler-Einstein four-manifold
is the product of two two-spheres. We also define the canonical two-form $J$ that exists on any $SE_7$. For $Q^{111}$
we have
\begin{align}
J=\tfrac{1}{8}(\mathrm{vol}_1+\mathrm{vol}_2+\mathrm{vol}_3)\,.
\end{align}

It is illuminating to rewrite the four-form flux appearing in our domain wall solutions using this basis:
\begin{align}\label{fexpression}
F=&
Zdt\wedge d\rho\wedge dx_{1}\wedge dx_{2}+H^{(1)}_2\wedge\omega^{(1)}+H^{(2)}_2\wedge \omega^{(2)}+H_2\wedge J \nn
&+df\wedge\eta\wedge\omega^{(1)}
+\frac{2}{3}f(J-\omega^{(2)})\wedge \omega ^{(1)}\,,
\end{align}
where
\begin{align}\label{Hexp}
H^{(1)}_2&=\lambda dx_1\wedge dx_2\,,\nn
H^{(2)}_2&= \tfrac{1}{3}e^{A-2B-V}\left[\beta(e^{-2U_3}+2e^{-4U_1+2U_3})+2\lambda f e^{-4U_1+2U_3}\right]dt\wedge d\rho\,,   \nn
H_2&=\tfrac{1}{3}e^{A-2B-V}\left[2\beta(e^{-2U_3}-e^{-4U_1+2U_3})-2\lambda f e^{-4U_1+2U_3}\right]dt\wedge d\rho \,.
\end{align}
The two-forms $H_2^{(1)}\equiv dB_1^{(1)}$ and $H_2^{(2)}\equiv dB_1^{(2)}$ are the field strengths of the two vector fields that lie in the two Betti vector multiplets. As discussed in 
\cite{Witten:2003ya,Benishti:2010jn,Franco:2009sp} 
there is a choice of $AdS_4$ boundary conditions for these vector fields corresponding to different
boundary CFTs. Essentially, this choice amounts to an electric-magnetic duality in the bulk.
For definiteness, we continue the discussion assuming standard boundary conditions for $B_1^{(1)}$ and $B_1^{(2)}$, which means that suitably wrapped membranes and fivebranes carry electric and magnetic baryonic charges, respectively. Other boundary conditions can be treated, {\it mutatis mutandis}, similarly.
From \eqref{Hexp} and the expansion \eqref{uvexpsads4}, on the one hand
we see that $\lambda$ is parametrising a deformation of the SCFT by a magnetic charge with respect to $B^{(1)}$. On the
other hand, we see that $\beta$ is parametrising electric charge with respect to $B^{(2)}$. In other words, the domain wall solutions we have constructed describe the dual
field theory at finite charge density with respect to one baryonic $U(1)$ when held in a finite magnetic field with respect to the other baryonic $U(1)$.

We next consider the field strength $H_2\equiv dB_1$ appearing in \eqref{Hexp}.
The consistent KK truncation \cite{Gauntlett:2009zw} is convenient for analysing this mode.
Observe that $H_2$ enters the ansatz for the four-form in precisely the same way that a two-form (also labelled $H_2$)
entered the consistent KK ansatz of \cite{Gauntlett:2009zw}. In \cite{Gauntlett:2009zw} it was discussed how the vector field 
$B_1$ mixes with another vector field $A_1$ that appears in the KK ansatz via
\begin{align}\label{akk}
\eta=\frac{1}{4}(d\psi+P_1+P_2+P_3+A_1)\,.
\end{align}
The mixing\footnote{That the $R$symmetry gauge-field comes from a mixing of the metric and the four-form was
noticed long ago e.g. in \cite{Fabbri:1999mk}.}
of $B_1$ and $A_1$ leads to a massless vector, dual to the $R$-symmetry current, 
and a massive vector with dimension $\Delta=5$, in the long vector multiplet of $OSp(2|4)$ mentioned above. 
From \eqref{Hexp} and \eqref{uvexpsads4} we have $H_2\sim e^{-4\rho}dt\wedge d\rho$ as $\rho\to\infty$ and hence it
is clear that the domain walls (in this section) do not carry
any electric or magnetic charge with respect to the $R$-symmetry $U(1)$.

In addition to the baryonic electric and magnetic charges carried by the the domain wall solutions, $D=4$ scalar fields are also active, corresponding to the constants $c_i$.
From our ansatz \eqref{eq:11d_ansatz} and from the expansion
\eqref{uvexpsads4} with \eqref{uvirvalues}
we can easily deduce that $c_1$ parametrises the expectation value of the scalar operator of dimension
$\Delta=4$ in the long vector multiplet of $OSp(2|4)$, mentioned above.  
The expansion \eqref{uvexpsads4} also implies that $c_2$ and $c_3$ correspond to deformations by, and expectation values for, 
the scalar and pseudo-scalar operators, respectively, that arise from the Betti vector multiplets. 

Observe that, as usual, in order to have a supersymmetric solution, flowing from a deformed $AdS_4$ to 
$AdS_2\times \mathbb{R}^2$, some fine tuning is required. Specifically, we chose $\beta =(2/3)^{1/2}\lambda$, 
which thus relates the value of the
electric and magnetic baryonic charges. Furthermore, the expansion \eqref{uvexpsads4}
and the specific values for the $c_i$ given in \eqref{uvirvalues}
imply that the deformations and expectation values of the scalar operators are also tuned.

To conclude this subsection, we comment on how the ansatz \eqref{eq:11d_ansatz} we have used to construct the supersymmetric domain walls
differs from the ansatz considered in \cite{Herzog:2009gd} and \cite{Klebanov:2010tj}. Indeed the expression for the four-form in \eqref{fexpression} makes
this straightforward. On the one hand if we set $f=\beta=0$ and $\lambda\ne 0$ in \eqref{eq:11d_ansatz}
then the ansatz is reduces to that considered in \cite{Herzog:2009gd}. On the other hand if we set 
$\lambda=f=0$ and $\beta\ne 0$ the ansatz is included in the ansatz of \cite{Klebanov:2010tj}
where {\it non-supersymmetric} $AdS_2$ solutions were found. The supersymmetric domain wall solutions we 
have found have $f,\beta,\lambda\ne 0$.

\section{Flows from $AdS_4$ to $AdS_2\times S^2$ and $AdS_2\times H^2$}
It is reasonably straightforward to generalise the supersymmetric black brane solutions which
interpolate from $AdS_4$ to $AdS_2\times \mathbb{R}^2$, that we have just constructed, to supersymmetric 
solutions that interpolate from $AdS_4$ to $AdS_2\times S^2$. It is also possible to exchange the $S^2$ factor with $H^2$
or $H^2/\Gamma$, where $\Gamma$ is a discrete group of isometries. We can treat all cases together by 
defining the two-dimensional metric
\begin{align}
ds_{4}^{2}=&\frac{dx^{2}}{1-k x^{2}}+\left(1-k x^{2}\right)\,d\phi_{4}^{2},\qquad J_{4}=dx\wedge d\phi_{4}\,,
\end{align}
with $k=1$ corresponding to the $S^2$ case, $k=-1$ corresponding to the $H^2$ case and $k=0$ corresponding to the $\mathbb{R}^2$ case already discussed.  It is also convenient to define a potential $P_4=xd\phi_4$ satisfying 
\begin{align}
dP_4=J_4\,.
\end{align}
One should be careful to note that when $k=+1$ the normalisation of the two-sphere metric $ds^2_4$ is not the same as those of $ds^2_i$ in \eqref{spherenorm}.

We then consider the ansatz for the $D=11$ metric and four-form given by
\begin{align}\label{eq:11d_ansatzv2}
ds^{2}=& -e^{2A}dt^{2}+e^{2B}\,ds_{4}^2+d\rho^{2}+e^{2U_{1}}\,\left(ds_{1}^{2}+ds_{2}^{2} \right)+e^{2U_{3}}\,ds_{3}^{2}+e^{2V}\,\eta^{2}\,,\notag\\
F=&dt\wedge d\rho\wedge\left(Z J_{4}+g_{1}J_{1}+g_{1}J_{2}+g_{3}J_{3} \right)+d\left[f \eta\wedge\left(J_{1}-J_{2}\right)\right]+\lambda\,J_{4}\wedge\left(J_{1}-J_{2} \right)\notag\\
=&dt\wedge d\rho\wedge\left(Z J_{4} +g_{1}J_{1}+g_{1}J_{2}+g_{3}J_{3}\right)\notag\\
&+f^{\prime}\,d\rho\wedge\eta\wedge\left(J_{1}-J_{2}\right)
+\left(2fJ_{3}+(\frac{1}{4}khf+\lambda)J_{4}\right)\wedge\left(J_{1}-J_{2} \right)\,,
\end{align}
where, as before,
\begin{align}
ds_{i}^{2}=&\tfrac{1}{8}\,\left(d\theta_{i}^{2}+\sin^{2}\theta_{i}\,d\phi_{i}^{2} \right),\quad \qquad J_{i}=\tfrac{1}{8}\,\sin\theta_{i}\,d\theta_{i}\wedge d\phi_{i},\qquad i=1,2,3\,.
\end{align}
A key new feature is that now
\begin{align}
\eta=&\tfrac{1}{4}\left(d\psi+P_1+P_2+P_3 + k hP_4\right),
\end{align}
and hence $d\eta=2\,\left(J_{1}+J_{2}+J_{3}\right)+\tfrac{1}{4}khJ_{4}$,
where $h$ is a constant to be determined. As before the functions $A,B,U_1,U_3,g_1,g_3,Z,f$ 
depend on the radial coordinate $\rho$ only. 
Again
$\lambda$ is a constant but note that it is only in the $k=0$ case that it can be scaled by scaling the coordinates $x,\phi_4$ and shifting $B$.

The ansatz satisfies the four-form Bianchi identity by construction. We also need to impose the four-form equation of motion.
Defining $G=e^{A+2B+V+4U_{1}+2U_{3}}$ we have
\begin{align}
\ast F=&-Ge^{-2A}\,\eta\wedge\left(g_{1}e^{-4U_{1}}J_{2}J_{3}J_{4}+g_{1}e^{-4U_{1}}J_{1}J_{3}J_{4}+g_{3}e^{-4U_{3}}J_{1}J_{2}J_{4}+Ze^{-4B}J_{1}J_{2}J_{3} \right)\notag\\
&+f^{\prime}Ge^{-2V-4U_{1}}\,dt\wedge J_{3}\wedge J_{4}\wedge\left(J_{2}-J_{1}\right)\notag\\
&+Ge^{-4U_{1}}\,dt\wedge d\rho\wedge\eta\wedge\left(2fe^{-4U_{3}}J_{4}+(\frac{khf}{4}+\lambda)e^{-4B}J_{3} \right)\left(J_{2}-J_{1}\right)\,.
\end{align}
The four-form equation of motion \eqref{eq:4form_eom} yields the relations
\begin{align}\label{eq:4form_solv2}
g_{1}=&\beta\,e^{A-2B-V-2U_{3}}\,,\notag\\
g_{3}=&-2\left(\beta+\lambda\,f+\frac{1}{8}kh(3+f^{2})\right)\,e^{A-2B-V-4U_{1}+2U_{3}}\,,\nn
Z=&\left(6-2f^{2}\right)\,e^{A+2B-V-4U_{1}-2U_{3}}\,,
\end{align}
where we have fixed a constant in $Z$,
and also the second order equation of motion
\begin{align}\label{eq:f_so_eomv2one}
-\left(Ge^{-2V-4U_{1}} f^{\prime}\right)^{\prime}+4fGe^{-4U_{1}}\left(e^{-4U_{3}}+\frac{k^2h^2}{64}e^{-4B}\right)+\frac{1}{4}\lambda khGe^{-4B-4U_{1}}\nn
=2fZ+ g_{3}\left(\lambda+\frac{1}{4}khf\right)\,.
\end{align}

\subsection{The supersymmetric flow equations}

To analyse the Killing spinor equation we define the elfbein
\begin{align}\label{genfram}
e^{0}&=e^{A}\,dt,\qquad \qquad e^{\bar i}=e^{B}\,E^{\bar i},\quad \bar i=1,2, \qquad\qquad
e^{3}=d\rho,\notag\\
 e^{\tilde{i}}&=e^{U_{1}}E^{\tilde{i}},\quad \tilde{i}=4,5,6,7\qquad\qquad
e^{\hat{i}}=e^{U_{2}}E^{\hat{i}},\quad \hat{i}=8,9,\qquad
 e^{\sharp}=e^{V}\,\eta,
\end{align}
where $E^{\bar i}$ is a frame for the metric on $S^2,\mathbb{R}^2$ or $H^2$ for $k=1,0,-1$, respectively, with $E^{\bar i}E^{\bar i}=ds^2_4$,  and 
$(E^{\tilde i},E^{\hat i})$ is a frame for $S^2\times S^2\times S^2$ with
$E^{\tilde i}E^{\tilde i}=ds^2_1+ds^2_2$ and $E^{\hat i}E^{\hat i}=ds^2_3$.
The Killing spinor equations with respect to this frame are written out in appendix \ref{killspinan}.
Here we just summarise how we can obtain flow equations preserving (generically) two supersymmetries, highlighting a feature
not present in the $k=0$ case.

We let $\hat\nabla$ be the spin connection for the metric with $A=B=U_1=U_3=V=0$. We take $\epsilon=e^{A/2}\epsilon_0$ and 
impose the projections \eqref{eq:projections} and \eqref{projs2} on $\epsilon_0$. We also impose the conditions associated with $Q^{111}$:
\begin{align}\label{eq:Q_condV2}
\hat{\nabla}_{\tilde{i}}\eta-\frac{e^{-U_{1}}}{2}\Gamma^{3}\Gamma_{\tilde{i}}\varepsilon_0=0\,,\notag\\
\hat{\nabla}_{\hat{i}}\varepsilon_0-\frac{e^{-U_{3}}}{2}\Gamma^{3}\Gamma_{\hat{i}}\varepsilon_0=0\,,\nn
\hat{\nabla}_{10}\varepsilon_0-\frac{e^{-V}}{2}\Gamma^{3}\Gamma_{10}\varepsilon_0=0\,,
\end{align}
and we note that the indices are tangent space indices with respect to the frame \eqref{genfram}.
We also impose the conditions
\begin{align}
(\hat{\nabla}_{\bar i}
+\frac{3h}{8}(P_{4})_{\bar i}\Gamma^{12})\varepsilon_0=0\,.
\end{align}
Recalling that we chose $P_4=xd\phi_4$ and taking into account the basis of vector fields dual to the frame \eqref{genfram}, 
in a coordinate basis we have
\begin{align}
\partial_{x}\varepsilon_0=0,\qquad 
(\partial_{\phi_4}+\frac{kx}{2}(1-h)\Gamma^{12})\varepsilon_0=0\,,
\end{align}
which is solved by taking 
\begin{align}
h=1\,,
\end{align} and $\varepsilon_0$ to be independent of $x,\phi_4$. This aspect of the preservation of supersymmetry is precisely
the same as the way in which it is preserved for supersymmetric wrapped branes \cite{Maldacena:2000mw}. 
Notice, in particular, that we can replace $H^2$ with $H^2/\Gamma$ to obtain a compact Riemann surface while still preserving supersymmetry.

The Killing spinor equations now reduce to the following first order equations
\begin{align}\label{eq:fo_systemv2}
A^{\prime}-\frac{e^{-A}}{3}\,\left(Ze^{-2B}+2g_{1}e^{-2U_{1}}+g_{3}e^{-2U_{3}} \right)&=0\notag\\
B^{\prime}-\frac{e^{-A}}{6}\,\left(2Ze^{-2B}-2g_{1}e^{-2U_{1}}-g_{3}e^{-2U_{3}} \right)-\frac{k}{8}e^{-2B+V}&=0\nn
U_{1}^{\prime}-e^{V-2U_{1}}+\frac{e^{-A}}{6}\left(Ze^{-2B}-g_{1}e^{-2U_{1}}+g_{3}e^{-2U_{3}}\right)&=0\notag\\
U_{3}^{\prime}-e^{V-2U_{3}}+\frac{e^{-A}}{6}\left(Ze^{-2B}+2g_{1}e^{-2U_{1}}-2g_{3}e^{-2U_{3}} \right)&=0\notag\\
V^{\prime}+2e^{V-2U_{1}}+e^{V-2U_3}
-4e^{-V}
+\frac{e^{-A}}{6}\left(Ze^{-2B}+2g_{1}e^{-2U_{1}}+g_{3}e^{-2U_{3}}  \right)+\frac{k}{8}e^{-2B+V}&=0\notag\\
f^{\prime}+2e^{V-2U_{3}}f+(\lambda+\frac{kf}{4})e^{V-2B}&=0
\end{align}
One can check that the second order equation \eqref{eq:f_so_eomv2one} is trivially satisfied. We also see that when $k=0$ the equations reduce
to those presented in \eqref{eq:fo_system}.

\subsection{$AdS_2\times S^2$ and $AdS_2\times H^2$ fixed points}
The flow equations \eqref{eq:fo_systemv2}
admit a two parameter family of supersymmetric $AdS_2\times S^2$ and $AdS_2\times H^2$ fixed points, which can be thought of 
as being parametrised by $\lambda,\beta$.
These comprise a sub-family of the supersymmetric solutions presented in section (3.2) of \cite{Donos:2008ug}.
We shall present two simple representative cases here, treating the others in appendix \ref{fptscurved}.

For $k=+1$ we have the $AdS_2\times S^2$ solution 
\begin{align}\label{eq:simple_S2}
& A=2^{7/3}\cdot 3^{1/6}\rho,\quad B=-\tfrac{1}{6}\,\ln 768,\quad \beta=1/4,\quad \lambda=1/\sqrt{2},\nn
&U_{1}= U_{3}=-\tfrac{1}{6}\,\ln \tfrac{3}{2},\quad V=-\tfrac{1}{6}\,\ln 12,\quad f=-\sqrt{2}.
\end{align}
This corresponds to the $D=11$ solution
\begin{align}\label{fptsols2}
ds^{2}&=\frac{1}{a^{2}}\,\left[ds^{2}\left(AdS_{2}\right)+\left(d\psi +P\right)^{2}\right]+\frac{a}{l_1}\,\left[d\tilde s_{1}^{2}+d\tilde s_{2}^{2}+d\tilde s_{3}^{2}
+d\tilde s_{4}^{2} \right]\,,\nn
F&=\mathrm{vol}(AdS_2)\wedge (\frac{3}{4l_1})(\tilde{\mathrm{vol}}_{1}+\tilde{\mathrm{vol}}_{2}+\tilde{\mathrm{vol}}_{3}+\tilde{\mathrm{vol}}_{4})\nn
&\qquad\qquad\qquad\qquad\qquad\qquad-\frac{1}{16\sqrt{2}}(\tilde{\mathrm{vol}}_{1}-\tilde{\mathrm{vol}}_{2})\wedge (\tilde{\mathrm{vol}}_{3}-\tilde{\mathrm{vol}}_{3})\,,
\end{align}
where $a=2^{7/3}\cdot 3^{1/6}$, $l_1=32\sqrt{3}$,
$d\tilde s_{i}^{2}$ being metrics of unit radius 2-spheres with volume-forms $\tilde{\mathrm{vol}}_{i}$ and
$dP=\tilde{\mathrm{vol}}_{1}+\tilde{\mathrm{vol}}_{2}+\tilde{\mathrm{vol}}_{3}+\tilde{\mathrm{vol}}_{4}$. Also, $ds^2(AdS_2)$, $\mathrm{vol}(AdS_2)$ are the metric and volume-form
for a unit radius $AdS_2$. This agrees with \cite{Donos:2008ug} (see footnote 3). Note that the manifold in this solution has topology
$AdS_2\times S^2\times S^2\times S^2\times S^3$.

For $k=-1$ we have the $AdS_2\times H^2$ solution
\begin{align}\label{eq:univ_ads2}
&A=4\rho,\quad  B=-\frac{3}{2}\,\ln2, \quad \beta=1/4,\quad  \lambda=0,\nn
&U_{1}=0,\quad
U_{3}=0,\quad V=0,\quad f=0\,.
\end{align}
This solution lies within the consistent KK truncation on an arbitrary $SE_7$ manifold presented in \cite{Gauntlett:2009zw} and
its existence was pointed out in section 6.1 of \cite{Gauntlett:2006qw}.

\subsection{Supersymmetric domain wall solutions}
We will construct domain wall solutions solving the flow equations \eqref{eq:fo_systemv2} that interpolate between
a deformation of $AdS_4$ in the UV and an $AdS_2\times S^2$ or $AdS_2\times H^2$ solution in the IR. For illustration, we only
discuss in detail the flows to the two IR $AdS_2$ solutions given in \eqref{eq:simple_S2} and \eqref{eq:univ_ads2}.
In the latter case, we provide the solution in closed form.

We saw in the last section when $k=0$ that $AdS_4$ is a solution to the flow equations \eqref{eq:fo_systemv2}. 
This is no longer the case when $k=\pm 1$. Instead we will look for domain wall solutions that asymptote to $AdS_4$ in the UV in Poincar\'e-type coordinates with the three-dimensional slices at constant $\rho$ having topology $\mathbb{R}\times S^2$ or $\mathbb{R}\times H^2$, respectively. 
This is precisely what happens in wrapped brane solutions \cite{Maldacena:2000mw}. The dual SCFT is now living on $\mathbb{R}\times S^2$ or $\mathbb{R}\times H^2$ with appropriate background $R$-symmetry currents switched on, in addition to the baryonic charges.

We can see this in more detail by generalising the discussion in
section \ref{sec:dual}. Writing the four-form flux in a similar way
to \eqref{fexpression},  in a KK reduction on $Q^{111}$ we get the $D=4$ field strengths
\begin{align}\label{Hexp2}
H^{(1)}_2&=(\lambda +\frac{kf}{4})J_4\,,\nn
H^{(2)}_2&= \tfrac{1}{3}e^{A-2B-V}\left[\beta(e^{-2U_3}+2e^{-4U_1+2U_3})+2(\lambda f+\frac{k}{8}(3+f^2)) e^{-4U_1+2U_3}\right]dt\wedge d\rho\,,   \nn
H_2&=\tfrac{1}{3}e^{A-2B-V}\left[2\beta(e^{-2U_3}-e^{-4U_1+2U_3})-2(\lambda f+\frac{k}{8}(3+f^2)) e^{-4U_1+2U_3}\right]dt\wedge d\rho \,.
\end{align}
With the fall-offs for the various functions that we find in the domain wall solutions presented below, $H^{(1)}$ and $H^{(2)}$ again correspond to magnetic and electric baryonic
charges, respectively. A new feature is that the field strength $F_2=dA_1$ for the gauge field appearing in the KK reduction via \eqref{akk} is given by
\begin{align}
F_2=kJ_4\,,
\end{align}
and, with the mixing with $H_2$, this gives rise to magnetic-charge with respect to the $R$-symmetry.

In the supersymmetric solutions of type IIB and $D=11$ supergravity describing D3-branes, M2-branes or M5-branes wrapping Riemann surfaces
just utilising $R$-symmetry currents \cite{Maldacena:2000mw,Gauntlett:2001qs}, it was found that only $AdS_2\times H^2/\Gamma$ fixed points are allowed i.e.
the genus of the Riemann surface is greater than one. The solutions that we construct here allow for $AdS_2\times S^2$ fixed points because of the presence of the
baryonic charges.

\subsubsection{Flow to $AdS_2\times S^2$}
To construct the supersymmetric flow from $AdS_4$ to the $AdS_2\times S^2$ solution \eqref{eq:simple_S2} we set $k=+1$,
$\beta=1/4$, $\lambda=1/\sqrt{2}$
and then develop expansions in the UV and IR.
The UV expansion is again governed by three constants, $c_i$, corresponding to the three relevant modes \eqref{modesone}.
In detail we have
\begin{align}\label{adsdrip}
A&=2 \rho+\frac{1}{80} \left(5-48 c_{2}^2+4 c_{3}^2\right) e^{-4 \rho}-\frac{2}{27} \left(12 c_{2}^3-\sqrt{2} c_{3}+c_{2} \left(3-5 c_{3}^2\right)\right) e^{-6 \rho}+\cdots\notag\\
B&=2 \rho+\frac{1}{80} \left(-5-48 c_{2}^2+4 c_{3}^2\right) e^{-4 \rho}+\frac{1}{108} \left(-96 c_{2}^3-\sqrt{2} c_{3}+10 c_{2} \left(3+4 c_{3}^2\right)\right) e^{-6 \rho}+\cdots\nn
U_{1}&=c_{2} e^{-2 \rho}+\left(-\frac{1}{8}+\frac{7 c_{2}^2}{5}-\frac{c_{3}^2}{30}\right) e^{-4 \rho}+\cdots+
c_{1} e^{-8 \rho}+\cdots\notag\\
U_{3}&=-2 c_{2} e^{-2 \rho}+\frac{1}{60} \left(15-96 c_{2}^2-2 c_{3}^2\right) e^{-4 \rho}+\cdots+
(c_1+\cdots)e^{-8\rho}\notag\\
V&=\left(-\frac{18 c_{2}^2}{5}-\frac{c_{3}^2}{30}\right) e^{-4 \rho}+\cdots
+(-6c_1+\cdots)e^{-8\rho}+\cdots\nn
f&=c_{3} e^{-2 \rho}+\frac{1}{4} \left(\sqrt{2}+16 c_{2} c_{3}\right) e^{-4 \rho}+\left(\frac{c_{2}}{\sqrt{2}}-\frac{3 c_{3}}{16}+\frac{59 c_{2}^2 c_{3}}{5}+\frac{c_{3}^3}{60}\right) e^{-6 \rho}+\cdots
\end{align}
where, for simplicity, we have suppressed a few order $e^{-6\rho}$ terms, but have included all terms at order $e^{-8\rho}$ that are linear in the
$c_i$.

Next we discuss the IR behaviour of the domain wall solutions flowing to \eqref{eq:simple_S2}. The $AdS_2$ fixed point has two irrelevant operators with
dimensions $\Delta=2$ and $\Delta=\sqrt{7}/2$ leading to an IR expansion depending on three constants, $d_i$, schematically given by 
\begin{align}\label{adsdrop}
A&=a \rho+d_{3}-8 e^{a\rho} d_{1}
+\frac{242}{7} e^{2a \rho} d_{1}^2+\cdots\notag\\
B&=-\frac{\log(768)}{6}+e^{a \rho} d_{1}+e^{(\sqrt{7}/2)a\rho} d_{2}+\cdots\nn
U_{1}&=-\frac{1}{6} \log\left(\frac{3}{2}\right)+e^{a\rho} d_{1}
-\frac{31}{7} e^{2a \rho} d_{1}^2+\cdots\notag\\
U_{3}&=-\frac{1}{6} \log\left(\frac{3}{2}\right)+e^{a \rho} d_{1}-e^{(\sqrt{7}/2)a\rho} d_{2}\cdots\notag\\
V&=-\frac{\log(12)}{6}+2 e^{a \rho} d_{1}
-\frac{74}{7} e^{2a \rho} d_{1}^2+\cdots\notag\\
f&=-\sqrt{2}+\frac{1}{3} \sqrt{2} \left(-1+\sqrt{7}\right) e^{(\sqrt{7}/2)a\rho} d_{2}
+\frac{1}{3} \sqrt{2} \left(-13+\sqrt{7}\right) e^{\left(1+\frac{\sqrt{7}}{2}\right) a\rho} d_{1} d_{2}+\cdots\,,
\end{align}
where we have included all $d_1 d_2$ terms.

\begin{figure}[t!]
\centering
\includegraphics[width=0.45\textwidth]{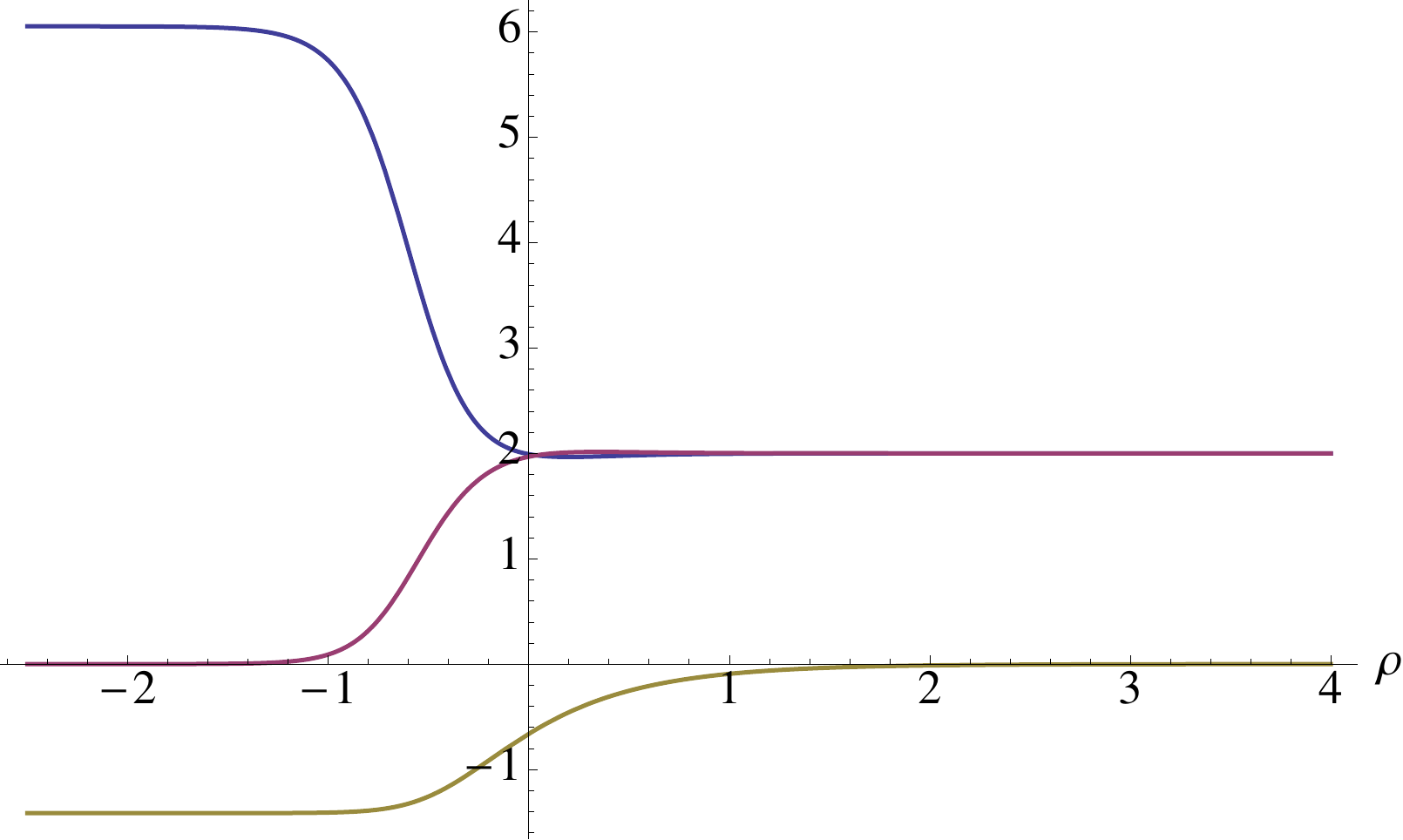}\qquad
\includegraphics[width=0.45\textwidth]{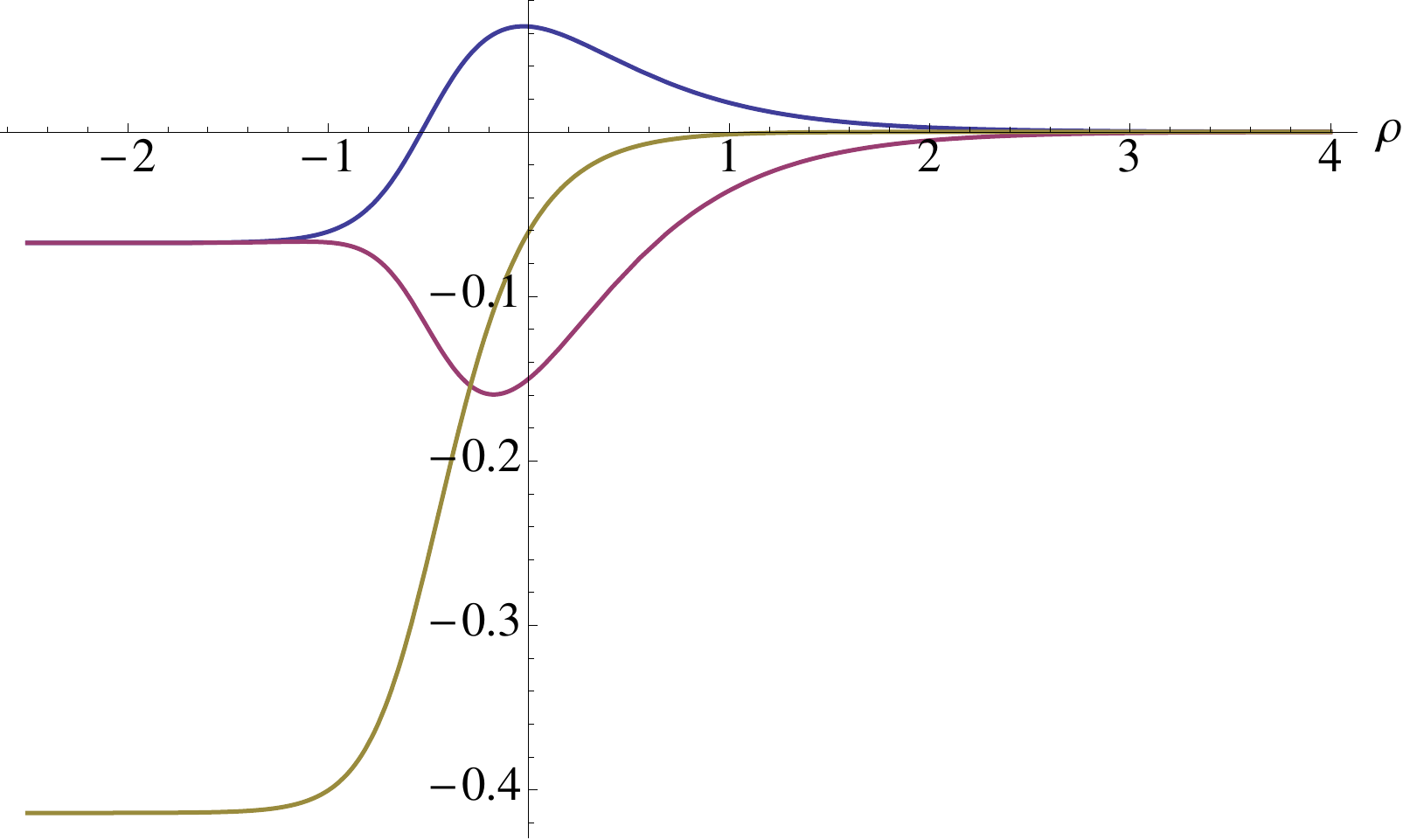}
\caption{Supersymmetric domain wall solutions interpolating between $AdS_4\times Q^{1,1}$ and $AdS_2\times {S}^2\times S^2\times S^2\times S^3$ given
by \eqref{fptsols2}.
In the left panel, top to bottom, we have plotted the functions $A'$ (blue), $B'$ (red), $f$ (green). In the right panel we have plotted, top to bottom from the left,
$U_1$ (blue), $U_3$ (red) and $V$ (green).
\label{fig:dwall2}}
\end{figure}

Using a shooting method we find the unique solution to the system of BPS equations
\eqref{eq:fo_systemv2} with boundary conditions \eqref{adsdrip}, \eqref{adsdrop} is given by
\begin{align}
&c_{1}=3.02\ldots,\qquad 
c_{2}=21.09\ldots,\qquad
c_{3}=2.57\ldots,\notag\\
&d_{1}=-0.032\ldots,\qquad
d_{2}=0.14\ldots,\qquad
d_{3}=-0.68\ldots\,.
\end{align}
In figure \ref{fig:dwall2} we have plotted the behaviour of the functions appearing in the solution.

\subsubsection{Flow to $AdS_2\times H^2$}
Next we discuss the domain wall solutions flowing from $AdS_4$ to the $AdS_2\times H^2$ solution \eqref{eq:univ_ads2}.
In fact for this case we can provide the analytic solution. With $k=-1$ we set
\begin{align}
&\beta=1/4,\quad  \lambda=0,\quad \qquad
U_{1}=
U_{3}=V=f=0.
\end{align}
The BPS equations \eqref{eq:fo_systemv2} simplify considerably and we find that after employing the simple change of coordinates
\begin{align}
\rho=\frac{1}{4}\log(8r^2-1)\,,
\end{align}
they can be easily integrated to give the $D=11$ solution
\begin{align}
&ds^2=-\left(2r-\frac{1}{4r}\right)^2dt^2+\left(2r-\frac{1}{4r}\right)^{-2}dr^2+r^2ds^2(H^2)\nn
&\qquad\qquad\qquad\qquad\qquad\qquad+\tfrac{1}{8}\left(d\tilde s^2_1+d\tilde s^2_2+d\tilde s^2_3\right)
+\tfrac{1}{4}(d\psi+P-P_4)\,,\nn
&F=dt\wedge dr  \left(6r^2\mathrm{vol}(H^2)+\tfrac{1}{32 r^2}
(\tilde{\mathrm{vol_1}}+\tilde{\mathrm{vol_2}}+\tilde{\mathrm{vol_3}})\right)\,.
\end{align}
where $d\tilde s^2_i$ and $\tilde{\mathrm{vol_i}}$ are the metric and volume-form of a unit radius two-sphere,
$dP=\tilde{\mathrm{vol_1}}+\tilde{\mathrm{vol_2}}+\tilde{\mathrm{vol_3}}$ and $dP_4=\mathrm{vol}(H^2)$.
One should note the close similarity with the solution found in section 3.4 of \cite{Gauntlett:2001qs} corresponding
to M2-branes wrapping supersymmetric $H_2$ cycles in Calabi-Yau five-folds.

\section{Final comments}
We have constructed supersymmetric domain wall solutions that interpolate between a deformation of the $AdS_4\times Q^{111}$ solution
in the UV and particular $AdS_2\times \mathbb{R}^2$, $AdS_2\times {S}^2$ an $AdS_2\times {H}^2$ solutions in the IR, first found in \cite{Donos:2008ug}. 
The ansatz we have considered only contains a two-parameter subfamily of the five-parameter family of supersymmetric $AdS_2$ solutions
constructed in section (3.2) of \cite{Donos:2008ug}. However, our ansatz can be generalised in a number of ways and it is likely that many if not all of
these $AdS_2$ solutions arise as IR fixed point of supersymmetric flows from $AdS_4\times Q^{111}$. For example, one can introduce
another metric function $U_2$ in \eqref{eq:11d_ansatzv2}
which will allow for additional relative squashing of the $S^2$ factors appearing in the metric. This will entail also
having another function $g_2$ in the four-form flux and in addition, generalising the function $f$ to four functions
via terms like $d(f_a\eta J_a)$. One can also generalise the constant $\lambda$ to 6 constants $\lambda_{ab}=\lambda_{(ab)}$
via $\lambda_{ab}J_a\wedge J_b$.

A more challenging task will be to find flow solutions that connect with the richer class of $AdS_2$ solutions of $D=11$ supergravity
that were constructed in section 5 of
\cite{Donos:2008ug}. For some of them we expect that this might be possible by flowing from $AdS_4\times SE_7$ solutions where the $SE_7$ manifold is of the type constructed
in \cite{Gauntlett:2004hh}.

We have noted that it is very likely that there is a consistent KK truncation of $D=11$ supergravity on $Q^{111}$, extending that of \cite{Gauntlett:2009zw} 
to include two additional $N=2$ Betti vector multiplets. Using the progress in constructing other supersymmetric $AdS_4$ black holes utilising
the special geometry of $N=2$ gauged supergravity coupled to vector multiplets \cite{Cacciatori:2009iz,Dall'Agata:2010gj,Hristov:2010ri,Barisch:2011ui},
it seems possible that some of 
the solutions we have found here using numerical methods, and generalisations thereof, can be found in closed form.

It would also be very interesting if further connections can be elucidated between the $D=11$ supergravity solutions that
we have constructed and the dual $N=2$ SCFTs in $d=3$ discussed in \cite{Franco:2008um,Franco:2009sp,Klebanov:2010tj,Benishti:2010jn}.

It is very likely that our supersymmetric solutions comprise a locus of solutions in
a larger moduli space of solutions of $D=11$ supergravity flowing from $AdS_4\times Q^{111}$ in the UV to $AdS_2$ solutions 
in the IR, which generically do not preserve supersymmetry. This would be analogous to the
magnetically charged solutions found in \cite{Almuhairi:2011ws,Donos:2011pn} and \cite{Donos:2011pn,Almheiri:2011cb} in the context of the $AdS_5\times S^5$ and $AdS_4\times S^7$ solutions
of type IIB and $D=11$ supergravity, respectively. As in \cite{Almuhairi:2011ws,Donos:2011pn,Almheiri:2011cb} we anticipate that the supersymmetric solutions flowing
to $AdS_2\times\mathbb{R}^2$ solutions we have constructed here might well be quantum critical points separating
novel phases. As in \cite{Almuhairi:2011ws,Donos:2011pn} evidence for this could be obtained by showing that adjacent non-supersymmetric
$AdS_2\times\mathbb{R}^2$ IR solutions suffer from instabilities. From this perspective, to hit the {\it stable}, supersymmetric, 
quantum critical
point, one needs to tune various deformation parameters in the UV (with expectation values fixed to ensure that the solutions
are regular in the IR) which is similar to the tuning required to hit quantum critical points in real systems.

\subsection*{Acknowledgements}
We thank S. Cremonesi for helpful discussions and J. Maldacena for correspondence which provided a catalyst this work.
AD is supported by an EPSRC Postdoctoral Fellowship and JPG is supported by a Royal Society Wolfson Award.

\appendix

\section{Killing spinor analysis}\label{killspinan}
We consider the metric and flux ansatz
\begin{align}\label{eq:11d_ansatzv2ap}
ds^{2}=& -e^{2A}dt^{2}+e^{2B}\,ds_{4}^2+d\rho^{2}+e^{2U_{1}}\,\left(ds_{1}^{2}+ds_{2}^{2} \right)+e^{2U_{3}}\,ds_{3}^{2}+e^{2V}\,\eta^{2}\notag\\
F
=&dt\wedge d\rho\wedge\left(Z J_{4} +g_{1}J_{1}+g_{1}J_{2}+g_{3}J_{3}\right)\notag\\
&+f^{\prime}\,d\rho\wedge\eta\wedge\left(J_{1}-J_{2}\right)
+\left(2fJ_{3}+(\frac{1}{4}khf+\lambda)J_{4}\right)\wedge\left(J_{1}-J_{2} \right)
\end{align}
where 
\begin{align}
ds_{i}^{2}=&\frac{1}{8}\,\left(d\theta_{i}^{2}+\sin^{2}\theta_{i}\,d\phi_{i}^{2} \right),\qquad J_{i}=\frac{1}{8}\,\sin\theta_{i}\,d\theta_{i}\wedge d\phi_{i},\quad i=1,2,3\notag\\
ds_{4}^{2}=&\frac{dx^{2}}{1-kx^{2}}+\left(1-kx^{2}\right)\,d\phi_{4}^{2},\qquad
J_{4}=dx\wedge d\phi_{4}\notag\\
\eta=&\tfrac{1}{4}\left(d\psi+P_1+P_2+P_3 + k hP_4\right),\quad dP_i=8J_i,\quad dP_4=J_4
\end{align}
Observe that $k=\pm 1$ for the case of $S^{2}, H^2$, respectively, $k=0$ for the case of $\mathbb{R}^{2}$ and $\lambda,h$ are constants.
We introduce the frame
\begin{align}\label{genframap}
e^{0}&=e^{A}\,dt,\qquad \qquad e^{\bar i}=e^{B}\,E^{\bar i},\quad \bar i=1,2, \qquad\qquad
e^{3}=d\rho,\notag\\
 e^{\tilde{i}}&=e^{U_{1}}E^{\tilde{i}},\quad \tilde{i}=4,5,6,7\qquad\qquad
e^{\hat{i}}=e^{U_{2}}E^{\hat{i}},\quad \hat{i}=8,,\qquad
 e^{\sharp}=e^{V}\,\eta,
\end{align}
where $E^{\bar i}$ is a frame for the metric on $S^2,\mathbb{R}^2$ or $H^2$ for $k=1,0,-1$, respectively, with $E^{\bar i}E^{\bar i}=ds^2_4$,  and 
$(E^{\tilde i},E^{\hat i})$ is a frame for $S^2\times S^2\times S^2$, with
$E^{\tilde i}E^{\tilde i}=ds^2_1+ds^2_2$ and $E^{\hat i}E^{\hat i}=ds^2_3$.
With respect to this frame, the covariant derivative can be written as
\begin{align}
\nabla\varepsilon=&\left(d+\frac{1}{4}\,\Gamma_{AB}\omega^{AB}\right)\varepsilon\nn
=&\hat{\nabla}\varepsilon-\frac{1}{2}(\Gamma_{45}+\Gamma_{67})\left(e^{V-2U_{1}}-e^{-V}\right)\,e^{\sharp}\varepsilon
-\frac{1}{2}\Gamma_{89}\left(e^{V-2U_{3}}-e^{-V}\right)\,e^{\sharp}\varepsilon\notag\\
&+\frac{1}{2}\,\left(e^{V-2U_{1}}-e^{-U_{1}}\right)\Gamma_{\sharp}(\Gamma_4 e^5-\Gamma_5 e^4+\Gamma_6 e^7-\Gamma_7 e^6)\varepsilon
+\frac{1}{2}\,\left(e^{V-2U_{3}}-e^{-U_{3}}\right)\Gamma_{\sharp}(\Gamma_8 e^9-\Gamma_9 e^8)
\varepsilon\notag\\
&-\frac{1}{2}A^{\prime}\Gamma_{30}e^{0}\varepsilon
-\frac{1}{2}B^{\prime}\Gamma_{3{\bar i}}e^{\bar i}\varepsilon
-\frac{1}{2}U_{1}^{\prime} \Gamma_{3\tilde{i}}e^{\tilde{i}}\varepsilon-\frac{1}{2}U_{3}^{\prime} \Gamma_{3\hat{i}}e^{\hat{i}}\varepsilon
-\frac{1}{2}V^{\prime}\Gamma_{3\,\sharp}e^{\sharp}\varepsilon
\notag\\
&-\frac{kh}{16}\,e^{V-2B}\Gamma_{12}e^{\sharp}\varepsilon
+\frac{kh}{16}e^{V-2B}\,\Gamma_{\sharp}(\Gamma_1 e^2-\Gamma_2 e^1)\varepsilon
-\frac{kh}{8}P_{4}\,\left[\Gamma_{45}+\Gamma_{67}+\Gamma_{89}\right] \varepsilon
\end{align}
where $\hat{\nabla}$ refers to the spin connection for the metric with $A=B=V=U_{1}=U_{2}=h=0$ and 
all the gamma matrix indices are tangent frame indices.

We now write down the explicit Killing spinor equations \eqref{eq:gravitinovariation}. For the $0$ component we have
\begin{align}\label{zerocomp}
&\hat{\nabla}_{0}\varepsilon-\frac{1}{2}A^{\prime}\Gamma_{30}\,\varepsilon-\frac{e^{-A}}{6}\Gamma_{3}\,\left[Ze^{-2B}\Gamma^{12}
+g_{1}e^{-2U_{1}}(\Gamma^{45}+\Gamma^{67})+g_{3}e^{-2U_{3}}\Gamma^{89} \right]\varepsilon\nn
&+\frac{e^{-2U_1}}{12}\Gamma_0\left(f'e^{-V}\Gamma^{3 \sharp}
+2fe^{-2U_3}\Gamma^{89}+(\frac{khf}{4}+\lambda)e^{-2B}\Gamma^{12}\right)(\Gamma^{45}-\Gamma^{67})\varepsilon
=0\,.
\end{align}
For the $1,2$ components we have
\begin{align}\label{onetwocomp}
&\hat{\nabla}_{\bar j}\varepsilon-\frac{kh}{8}P_{4}{}_{\bar j}\,\left[\Gamma^{45}+\Gamma^{67}+\Gamma^{89}\right] \varepsilon
-\frac{kh}{16}e^{V-2B}\,\Gamma_{\bar i \sharp}J_{4}^{\bar i}{}_{\bar j}\varepsilon-\frac{1}{2}B^{\prime}\Gamma_{3\bar j}\,\varepsilon\notag\\
&+\frac{e^{-A}}{12}\Gamma_{\bar j}\Gamma^{03}\,\left[-2Ze^{-2B}\Gamma^{12}+g_{1}e^{-2U_{1}}(\Gamma^{45}+\Gamma^{67})+g_{3}e^{-2U_{3}}\Gamma^{89} \right]\varepsilon\nn
&+\frac{e^{-2U_1}}{12}\Gamma_{\bar j}\left(f'e^{-V}\Gamma^{3 \sharp}
+2fe^{-2U_3}\Gamma^{89}-2(\frac{khf}{4}+\lambda)e^{-2B}\Gamma^{12}\right)(\Gamma^{45}-\Gamma^{67})\varepsilon
=0\,,
\end{align}
where we defined $J_4=(1/2)J_{4\bar i\bar j}E^{\bar i} E^{\bar j}$.
For the $3$ component we have
\begin{align}\label{threecomp}
&\hat{\nabla}_{3}\,\varepsilon-\frac{e^{-A}}{6}\Gamma_0\left[Ze^{-2B}\Gamma^{12}+g_{1}e^{-2U_{1}}\Gamma^{45}+g_{1}e^{-2U_{1}}\Gamma^{67}+g_{3}e^{-2U_{3}}\Gamma^{89}  \right]\varepsilon=0\notag\\
&+\frac{e^{-2U_1}}{12}\Gamma_{3}\left(-2f'e^{-V}\Gamma^{3 \sharp}
+2fe^{-2U_3}\Gamma^{89}+(\frac{khf}{4}+\lambda)e^{-2B}\Gamma^{12}\right)(\Gamma^{45}-\Gamma^{67})\varepsilon=0\,.
\end{align}
For the $\tilde i= 4,5$ components we have
\begin{align}\label{fourfivecomp}
&\hat{\nabla}_{\tilde{i}}\varepsilon-\frac{1}{2}U_{1}^{\prime}\Gamma_{3\tilde{i}}\varepsilon-\frac{1}{2}\left(e^{V-2U_{1}}-e^{-U_{1}} \right)\Gamma_{\tilde{j}\,\sharp}J_{1}^{\tilde{j}}{}_{\tilde{i}}\varepsilon\notag\\
&+\frac{e^{-A}}{12}\Gamma_{\tilde j}\Gamma^{03}\,\left[Ze^{-2B}\Gamma^{12}+g_{1}e^{-2U_{1}}(-2\Gamma^{45}+\Gamma^{67})+g_{3}e^{-2U_{3}}\Gamma^{89} \right]\varepsilon\nn
&+\frac{e^{-2U_1}}{12}\Gamma_{\tilde j}\left(f'e^{-V}\Gamma^{3 \sharp}
+2fe^{-2U_3}\Gamma^{89}+(\frac{khf}{4}+\lambda)e^{-2B}\Gamma^{12}\right)(-2\Gamma^{45}-\Gamma^{67})\varepsilon
=0\,,
\end{align}
where we defined $J_1=(1/2)J_{1\tilde i\bar j}E^{\tilde i} E^{\tilde j}$. 
For the $\tilde i= 6,7$ components we have
\begin{align}\label{sixsevencomp}
&\hat{\nabla}_{\tilde{i}}\varepsilon-\frac{1}{2}U_{1}^{\prime}\Gamma_{3\tilde{i}}\varepsilon-\frac{1}{2}\left(e^{V-2U_{1}}-e^{-U_{1}} \right)\Gamma_{\tilde{j}\,\sharp}J_{2}^{\tilde{j}}{}_{\tilde{i}}\varepsilon\notag\\
&+\frac{e^{-A}}{12}\Gamma_{\tilde j}\Gamma^{03}\,\left[Ze^{-2B}\Gamma^{12}+g_{1}e^{-2U_{1}}(\Gamma^{45}-2\Gamma^{67})+g_{3}e^{-2U_{3}}\Gamma^{89} \right]\varepsilon\nn
&+\frac{e^{-2U_1}}{12}\Gamma_{\tilde j}\left(f'e^{-V}\Gamma^{3 \sharp}
+2fe^{-2U_3}\Gamma^{89}+(\frac{khf}{4}+\lambda)e^{-2B}\Gamma^{12}\right)(\Gamma^{45}+2\Gamma^{67})\varepsilon
=0\,,
\end{align}
where we defined $J_2=(1/2)J_{2\tilde i\bar j}E^{\tilde i} E^{\tilde j}$.
For the $8,9$ components we have
\begin{align}\label{eightninecomp}
&\hat{\nabla}_{\hat{i}}\varepsilon-\frac{1}{2}U_{3}^{\prime}\Gamma_{3\hat{i}}\varepsilon-\frac{1}{2}\left(e^{V-2U_{3}}-e^{-U_{3}} \right)\Gamma_{\hat{j}\,\sharp}J_{3}^{\hat{j}}{}_{\hat{i}}\varepsilon\notag\\
&+\frac{e^{-A}}{12}\Gamma_{\hat j}\Gamma^{03}\,\left[Ze^{-2B}\Gamma^{12}+g_{1}e^{-2U_{1}}(\Gamma^{45}+\Gamma^{67})-2g_{3}e^{-2U_{3}}\Gamma^{89} \right]\varepsilon\nn
&+\frac{e^{-2U_1}}{12}\Gamma_{\hat j}\left(f'e^{-V}\Gamma^{3 \sharp}
-4fe^{-2U_3}\Gamma^{89}+(\frac{khf}{4}+\lambda)e^{-2B}\Gamma^{12}\right)(\Gamma^{45}-\Gamma^{67})\varepsilon
=0\,,
\end{align}
where we defined $J_3=(1/2)J_{3\hat i\hat j}E^{\hat i} E^{\hat j}$.
Finally, for the $\sharp$ component we have
\begin{align}\label{tencomp}
&\hat{\nabla}_{\sharp}\varepsilon-\frac{kh}{16}\,e^{V-2B}\Gamma^{12}\varepsilon-\frac{1}{2}(\Gamma^{45}+\Gamma^{67})
\left( e^{V-2U_{1}}-e^{-V}\right)\varepsilon
-\frac{1}{2}\,\Gamma^{89}\left( e^{V-2U_{3}}-e^{-V}\right)\varepsilon\notag\\
&-\frac{1}{2}V^{\prime}\Gamma_{3\,\sharp}\varepsilon
+\frac{e^{-A}}{12}\Gamma_{\sharp}\Gamma^{03}\,\left[Ze^{-2B}\Gamma^{12}+g_{1}e^{-2U_{1}}(\Gamma^{45}+\Gamma^{67})+g_{3}e^{-2U_{3}}\Gamma^{89} \right]\varepsilon\nn
&+\frac{e^{-2U_1}}{12}\Gamma_{\sharp}\left(-2f'e^{-V}\Gamma^{3 \sharp}
+2fe^{-2U_3}\Gamma^{89}+(\frac{khf}{4}+\lambda)e^{-2B}\Gamma^{12}\right)(\Gamma^{45}-\Gamma^{67})\varepsilon
=0\,.
\end{align}

To obtain the supersymmetric domain wall flow equations, it is natural to impose the projections for the Poincar\'e supersymmetries of
the $AdS_4\times Q^{111}$ solution. Thus we impose
\begin{align}
&\Gamma^{4567}\varepsilon=-\varepsilon,\quad \Gamma^{4589}\varepsilon=-\varepsilon,\quad\Gamma^{453\sharp}\varepsilon=-\varepsilon
\qquad\Rightarrow\Gamma^{012}\,\varepsilon=-\varepsilon\,.
\end{align}
We also demand that
\begin{align}
&\Gamma^{1245}\varepsilon=-\varepsilon\,,
\end{align}
leading to an overall preservation of two supersymmetries.
By considering \eqref{zerocomp}, \eqref{threecomp}, we choose the Killing spinor to be of the form
\begin{align}
\varepsilon=e^{A/2}\varepsilon_0\,,
\end{align}
with $\varepsilon_0$ independent of the $t,\rho$ coordinates. 
Using the fact that $\hat{\nabla}$ is the connection on the product of the Sasaki-Einstein metric on $Q^{111}$ 
with the metric $-dt^{2}+ds_{4}^2+d\rho^{2}$, we can impose
\begin{align}\label{eq:Q_condV22}
\hat{\nabla}_{\tilde{i}}\varepsilon_0-\frac{e^{-U_{1}}}{2}\Gamma_{3\tilde{i}}\varepsilon_0=0\,,\notag\\
\hat{\nabla}_{\hat{i}}\varepsilon_0-\frac{e^{-U_{3}}}{2}\Gamma_{3\hat{i}}\varepsilon_0=0\,,\nn
\hat{\nabla}_{11}\varepsilon_0-\frac{e^{-V}}{2}\Gamma_{3\sharp}\varepsilon_0=0\,,
\end{align}
and we note that the indices are tangent space indices with respect to the frame \eqref{genframap}.
Finally, for the $1,2$ components we impose
\begin{align}
\hat{\nabla}_{\bar i}\varepsilon-\frac{3kh}{8}P_{4}{}_{\bar i} \Gamma^{12} \varepsilon=0\,.
\end{align}
This is essentially the same way in which supersymmetry is preserved for branes wrapping supersymmetric cycles \cite{Maldacena:2000mw}.
In particular, as we explain in the text, we should choose $h=1$ and then the Killing spinors
are independent of the coordinates on the $S^2,H^2$ factor when $k=\pm1$, respectively (they are trivially independent of the
coordinates on the $\mathbb{R}^2$ factor when $k=0$).

\section{A family of $AdS_{2}\times\mathbb{R}^2$ fixed point solutions}\label{sec:AdS_familyB}
The general one parameter family of supersymmetric $AdS_2\times \mathbb{R}^2$ solutions to the systems of equations
\eqref{eq:fo_system} can be expressed in terms of a free parameter $m$ via
\begin{align}\label{fullsetsol}
&A=a\rho,\qquad a=\frac{ 2^{4/3}}{3^{1/6}}\left(\frac{4m-3}{m-1} \right)^{1/3}\left(4m^{2}-3 \right)^{1/6}\,
\quad \beta=\left(2m-1 \right)\,\sqrt{\frac{3}{m}\frac{m-1}{4m^{2}-3}}\,\lambda\notag\\
&U_{1}=\frac{1}{6}\ln\left(\frac{3}{4}\frac{4m-3}{\left(m-1\right)\left(4m^{2}-3\right)}\right),\quad U_{3}=\frac{1}{6}\ln\left(6\frac{\left(4m-3 \right)\left(m-1\right)^{2}}{4m^{2}-3} \right)\notag\\
&V=\frac{1}{6}\ln\left(48\frac{\left(m-1\right)^{2}}{\left(4m^{2}-3\right)\left(4m-3\right)^{2}} \right),\quad f=-2\sqrt{3}\sqrt{\frac{m\,\left(m-1\right)}{4m^{2}-3}}\notag\\
&B=\frac{1}{12}\ln\left(\frac{\lambda^{6}}{3\cdot 2^{10}\,m^{3}}\left(m-1\right)\left(4m^{2}-3\right)\left(4m-3\right)^2\right)
\end{align}
with $m>1$.
The solution we considered in the text in \eqref{fptsol} is obtained by setting $m=3/2$.

The resulting $D=11$ solution can be written in the form
\begin{align}\label{uplift2ap}
ds^{2}=& \frac{1}{a^2}\left[ds^2(AdS_2)+a^3\left(\gamma(dx_{1}^{2}+dx_{2}^{2})
+\tfrac{1}{l_1}\left(d\tilde s_{1}^{2}+d\tilde s_{2}^{2} +d\tilde s_{3}^{2}\right)\right)+
(d\psi+P)^2\right]\,,\notag\\
F=&\frac{1}{a^3}\mathrm{vol}(AdS_2)\wedge \left[(2l_1+l_3)\gamma dx_{1}\wedge dx_{2}+\frac{l_1+l_3}{l_1}\left(\tilde{\mathrm{vol}}_1+\tilde{\mathrm{vol}}_2\right)+\frac{2l_1}{l_3}\tilde{\mathrm{vol}}_3\right]\nn
&+2m_{13}\left[-\gamma dx_1\wedge dx_2    +  \frac{1}{l_3}\tilde{\mathrm{vol}}_3\right] \wedge\frac{1}{l_1}(\tilde{\mathrm{vol}}_1-\tilde{\mathrm{vol_2}})
\end{align}
with $P=P_1+P_2+P_3$, $l_1=2^5(4m^2-3)^{1/2}/3^{1/2}$, $l_3=2^4(4m^2-3)^{1/2}/(3^{1/2}(m-1))$, $m_{13}=-l_1m^{1/2}/(2(m-1)^{1/2})$ and
$\gamma=\lambda((m-1)^{1/2}/(8m^{1/2})$. Furthermore,
$ds^2(AdS_2)$ and $\mathrm{vol}(AdS_2)$ are the metric and volume-form on a unit radius $AdS_2$.
This agrees with the solutions of section (3.2) in \cite{Donos:2008ug} (with $m_{12}=0$) up to the typos mentioned in footnote 3.

\section{$AdS_{2}\times S^2$ and $AdS_2\times H^2$ fixed point solutions}\label{fptscurved}
With $k=\pm 1$ we can construct a two-parameter family of $AdS_2$ solutions to the BPS equations
\eqref{eq:fo_systemv2} that are a sub-family of those in \cite{Donos:2008ug}. With $A=a\rho$ we can parametrise the solutions by
the constants $U_{1}$ and $U_{3}$:
\begin{align}\label{eq:ads2fpc}
a&=4\,\sqrt{3}e^{2U_{3}}\,\left( -3+2 e^{4 U_{1}+2 U_{3}}+4 e^{2 U_{1}+4 U_{3}}\right)^{-1/2}\,,\notag\\
e^{2V}&=16\,a^{-2}\,,\notag\\
e^{2B}&=-\frac{e^{2 \left(U_{1}+U_{3}\right)} \left(-3+2 e^{4 U_{1}+2 U_{3}}+4 e^{2 U_{1}+4U_{3}}\right) k}{8 \left(-3 e^{2 U_{1}}+8 e^{4 \left(U_{1}+U_{3}\right)}+2 e^{2 U_{1}+6 U_{3}}+2 e^{2 U_{3}} \left(-3+e^{6 U_{1}}\right)\right)}\,,\notag\\
f&=-\frac{1}{\sqrt{3}}\,\sqrt{9-e^{2 U_{1}-4 U_{3}} \left(e^{2 U_{1}}+2 e^{2 U_{3}}\right) \left(-3+2 e^{4 U_{1}+2 U_{3}}+4 e^{2 U_{1}+4 U_{3}}\right)}\,,\notag\\
\lambda&=-\left(2 e^{2B-2U_{3}}+\frac{k}{4}\right)\,f\,,\notag\\
\beta&=\frac{e^{2 U_{1}} \left(-3+2 e^{4 U_{1}+2 U_{3}}+4 e^{2 U_{1}+4 U_{3}}\right)   \left(-3+4 e^{3 \left(U_{1}+U_{3}\right)} \sinh\left(U_{1}-U_{3}\right)\right)}{12 \left(-3 e^{2 U_{1}}+8 e^{4 \left(U_{1}+U_{3}\right)}+2 e^{2 U_{1}+6 U_{3}}+2 e^{2 U_{3}} \left(-3+e^{6 U_{1}}\right)\right)}\,k\,.
\end{align}
The parameters $U_{1}$ and $U_{3}$ need to be constrained so that the above formulae give physically sensible answers i.e. they are all real numbers. 
One can check that the solutions \eqref{eq:simple_S2} and \eqref{eq:univ_ads2} are recovered by setting $U_{1}=U_{3}=-\tfrac{1}{6}\ln\tfrac{3}{2}$, $k=1$ and $U_{1}=U_{3}=0$, $k=-1$, respectively.

To compare with previous work that appeared in section 3.2 of \cite{Donos:2008ug} we need to make the identifications
(see footnote 3)
\begin{align}
&e^{A_{there}}=\frac{1}{a}\,,\quad l_{1}=l_{2}=8ae^{-2U_{1}}\,,\quad l_{3}=8ae^{-2U_{3}}\,,\quad l_{4}=k a e^{-2B}\,,\notag\\
&m_{13}=-m_{14}=\frac{l_{1}l_{3}f}{64},\quad m_{12}=0\,.
\end{align}

\end{document}